\documentclass[entropy,article,accept,moreauthors]{mdpi}

\usepackage{amsmath,amssymb}
\usepackage{xcolor}

\usepackage{textcomp}

\firstpage{1} 
\pubvolume{22}
\issuenum{3}
\articlenumber{286}
\pubyear{2020}
\copyrightyear{2020}
\history{Received: 04 February 2020; Accepted: 27 February 2020; Published:  29 February 2020}
\updates{yes} 

\Title{Entropy Dynamics of Phonon Quantum States Generated by Optical Excitation of a Two-Level~System}

\Author{Thilo Hahn, Daniel Wigger\orcidA{} and Tilmann Kuhn *\orcidB{}}

\AuthorNames{Thilo Hahn, Daniel Wigger and Tilmann Kuhn}

\address[1]{%
Institut f\"ur Festk\"opertheorie, Universit\"at M\"unster, Wilhelm-Klemm-Str. 10 48149 M\"unster, Germany; t.hahn@wwu.de (T.H.); d.wigger@wwu.de (D.W.)}

\corres{\hangafter=1 \hangindent=1.05em \hspace{-0.82em} Correspondence: tilmann.kuhn@uni-muenster.de}

\abstract{In quantum physics, two prototypical model systems stand out due to their wide range of applications. These are the two-level system (TLS) and the harmonic oscillator. The former is often an ideal model for confined charge or spin systems and the latter for lattice vibrations, i.e., phonons. Here, we couple these two systems, which leads to numerous fascinating physical phenomena. Practically, we consider different optical excitations and decay scenarios of a TLS, focusing on the generated dynamics of a single phonon mode that couples to the TLS. Special emphasis is placed on the entropy of the different parts of the system, predominantly the phonons. While,~without~any decay, the entire system is always in a pure state, resulting in a vanishing entropy, the complex interplay between the single parts results in non-vanishing respective entanglement entropies and non-trivial dynamics of them. Taking a decay of the TLS into account leads to a non-vanishing entropy of the full system and additional aspects in its dynamics. We demonstrate that all aspects of the entropy's behavior can be traced back to the purity of the states and are illustrated by phonon Wigner functions in phase space.}


\begin{document}
\section{Introduction}
Entropy is one of the most fundamental concepts in physics. According to the second law of thermodynamics, in~a closed system, it never decreases, which has far reaching consequences, from~the limited efficiency of thermodynamic machines~\cite{kondepudi2014mode} to cosmological implications~\cite{gunzig1987entr,la2012blac}. Under~thermal equilibrium conditions, it determines the state of a thermodynamic system: In a closed system, the~realized state is the one with maximal entropy; in a system in thermal contact with a heat bath, the~realized state results from an interplay between energy and entropy and is governed by the minimum of the free energy~\cite{toda1992equi}. Under~nonequilibrium conditions, the~second law of thermodynamics prohibits the decrease of the entropy of a closed system; however, this does not hold for the entropy of a subsystem which is interacting with other subsystems or with its surroundings~\cite{kubo2012stat,breuer2002theo}. In~this case, the~study of the dynamics of the entropy of these subsystems provides valuable information on the evolution of the nature of the system's state~\cite{eisert2010coll}. 

From the point of view of information science, entropy is closely related to the imperfect knowledge about a system~\cite{shannon1948math,liang2006info}. As~such, it plays a key role in all fields related to information processing and communication, and, in~particular, in~the highly topical fields of quantum information and communication, where the entropy is closely related to phenomena like purity of quantum states, entanglement, and~decoherence~\cite{nielsen2002quan}.

In this paper, we study the entropy dynamics in a prototypical model of quantum mechanics and quantum information theory. It consists of two subsystems, a~quantum-mechanical two-level system (i.e., a~representation of a qubit) which can be manipulated by an external field (e.g., a~light field) and which is coupled to a harmonic oscillator (e.g., a~single phonon mode or a nanomechanical oscillator). The~generation of specific quantum states of such a harmonic oscillator and the manipulation of these states has recently attracted much interest~\cite{hofheinz2009synt,o2010qua,reiter2011gene,satzinger2018quan}. Prominent examples are coherent states and Schr\"odinger cat states, i.e.,~superpositions of coherent states. We analyze the entropy dynamics of the two subsystems after excitation with a short optical pulse or a pair of such pulses. In~particular, we compare the case of a unitary evolution in the absence of damping processes, when the coupled system remains in a pure state, with~the case of a decaying two-level system resulting in a mixed state also of the combined system. We will show that the analysis of the time-dependent entropy provides interesting insight into the nature of the quantum state of the two~subsystems.

\section{Theory}
We consider a two-level system (TLS) which can be excited and de-excited by a resonant optical field ${\bf E}$. Additionally, a~pure dephasing coupling to a single phonon (ph) mode is taken into account. Thus, the~Hamiltonian reads~\cite{mahan1981many}
\begin{equation}
		H=\hbar\Omega \left| x \right>\left< x \right| - \big[ {\bf M}\cdot {\bf E}(t) \left| x \right>\left< g \right| + {\bf M}^\ast\cdot {\bf E}^\ast(t) \left| g \right>\left< x \right|   \big] + \hbar\omega_{\rm ph} \hat{b}^\dagger\hat{b} + \hbar g \big( \hat{b} + \hat{b}^\dagger \big)\left| x \right>\left< x \right|\ . \label{eq:H}
	\end{equation}

The states $\left| g \right>$ and $\left| x \right>$ describe ground and excited state of the TLS with an energy splitting of $\hbar\Omega$, respectively. The~time dependent optical driving is mediated by the dipole matrix element~${\bf M}$. Phonons~with the discrete energy $\hbar\omega_{\rm ph}$ are created and annihilated by $\hat{b}^\dagger$ and $\hat{b}$, respectively. For~simplicity, the~coupling constant of the exciton-phonon interaction $g$ is supposed to be~real.

Such a TLS system coupled to a single bosonic mode is a prototypical model that can be considered for the description of various solid state systems. For~the TLS, one might think of an exciton in a single semiconductor quantum dot~\cite{zrenner2002cohe} or excitations of defects in insulators, like diamond~\cite{aharonovich2011diam} or hexagonal boron nitride~\cite{wigger2019phon}, while the phonon could be an optical mode~\cite{roca1994pola}, a~local mode~\cite{gali2011ab}, a~van Hove singularity~\cite{debald2002cont}, or~the mechanical excitation of a microresonator~\cite{munsch2017reso}.

Phonon-induced transitions between the states $\left|g\right>$ and $\left|x\right>$ of the TLS are negligible because of the strong energy mismatch between the exciton energy, which is of the order of one or a few electronvolts (eV), while phonon energies range from a few micro-electronvolts (\textmu eV) (for micromechanical resonators) up to a few tens of milli-electronvolts (meV) (for optical phonon modes). The~linear coupling in the phonon displacement reflects typical electron-phonon interaction mechanisms in solids like deformation potential coupling, piezoelectric coupling, or~Fr\"ohlich coupling~\cite{ferry1991semiconductors}. Although~extensions of this model have been considered that take a quadratic coupling to the phonons into account~\cite{munn1978theo,muljarov2004deph,machnikowski2006chan,chenu2019two}, the~original independent boson model~\cite{duke1965phon} in Equation \eqref{eq:H} is successfully used in different contexts. It reproduces recent linear and nonlinear spectroscopy signals~\cite{stock2011acou,wigger2019phon}, Rabi~oscillations~\cite{wigger2018rabi}, and~rotations~\cite{ramsay2010phon} in excellent agreement with experiments, to~name just a~few.

The possible states of the entire system can be separated into the phonons forming product states with the ground state of the TLS and those forming product states with the excited state
\begin{equation}
		\left| g\right>\otimes \left| ph_{\rm g}\right> \qquad {\rm and}\qquad \left| x\right>\otimes \left| ph_{\rm x}\right>\ .
	\end{equation}

From the full density matrix of the system $\rho$, we can calculate the one of a subsystem by tracing over the respective other, i.e.,
\begin{equation}
		\rho_{\rm TLS} = {\rm Tr}_{\rm ph}(\rho) \qquad {\rm and}\qquad \rho_{\rm ph} = {\rm Tr}_{\rm TLS}(\rho)\ .
	\end{equation}

In the same way as in Reference~\cite{hahn2019infl}, we model a decay of the excited state with the rate $\Gamma$ via the Lindblad dissipator
\begin{equation}
		\mathcal D(\rho) = \Gamma \left[ \left|g\right>\left< x\right| \rho \left|x\right>\left< g\right| 
		-\frac12 \big\{ \left|x\right>\left< x\right|, \rho \big\}\right] ,
	\end{equation}
leading to the master equation for the density matrix:
\begin{equation}
		\frac{\rm d}{{\rm d}t}\rho = \frac{1}{i\hbar} [H,\rho] + \mathcal D(\rho) \ .
	\end{equation}

We will not take an additional phenomenological pure dephasing of the TLS into account. We~will study different regimes of decay rates $\Gamma$ compared to the characteristic phonon frequency $\omega_{\rm ph}$. On~the one hand, when describing, for~example, optical phonons with energies in the range of tens of meV the decay time of the TLS is typically much longer than a phonon period, i.e.,~$\Gamma\ll \omega_{\rm ph}$~\cite{wigger2019phon}. On~the other hand, when considering typical mechanical resonators with phonon energies in the \textmu eV range we have $\Gamma > \omega_{\rm ph}$~\cite{auffeves2014opti}.

As we have explained in Reference~\cite{hahn2019infl}, the~entire quantum state and especially the Wigner function of the phonons can be calculated analytically when considering a series of ultrafast laser pulses to drive the TLS. Especially if the pulse duration is much shorter than the phonon period, the~pulses can be approximated by delta-functions as
\begin{equation}
		\frac{{\bf M}\cdot {\bf E}(t)}{\hbar} = \sum_j\frac{\theta_j}{2} \exp \left[-i\left(\Omega -\frac{g^2}{\omega_{\rm ph}} \right)t +i\phi_j \right] \delta(t-t_j)\ .
	\end{equation}

The pulses excite the TLS at times $t_j$ with pulse areas $\theta_j$ and phases $\phi_j$. By~this choice and the introduction of the generating functions
	\begin{subequations}\label{eq:gen_func}
\begin{align}
		Y_\alpha(t) &= \Big< \left|g\right>\left< x\right| \exp(-\alpha^\ast \hat{b}^\dagger) \exp(\alpha \hat{b})  \Big> \ ,\\
		C_\alpha(t) &= \Big< \left|x\right>\left< x\right| \exp(-\alpha^\ast \hat{b}^\dagger) \exp(\alpha \hat{b})  \Big> \ ,\\
		F_\alpha(t) &= \Big<  \exp(-\alpha^\ast \hat{b}^\dagger) \exp(\alpha \hat{b})  \Big>\ ,
	\end{align}
	\end{subequations}
with $\left< \hat{A}\right> = {\rm Tr}\left(\rho \hat{A}\right)$ denoting the expectation value of an operator $\hat{A}$, a~closed system of partial differential equations for the time-evolution of the generating functions is obtained and all phonon assisted density matrices can be calculated analytically without approximations. Note that $F_\alpha$ contains the entire information on the phonon system, while $C_\alpha$ describes the phonon assisted occupation of the excited state, i.e.,~the phonons in $\left|x\right>\otimes\left|ph_{\rm x}\right>$, and~$Y_\alpha$ is the phonon assisted~coherence.

Our analysis of the phonon quantum states is based on their Wigner function~\cite{schleich2011quan}
\begin{equation}
		W(U,\Pi) = \frac{1}{4\pi}\int\limits_{-\infty}^\infty \left<U+\frac{X}{2}\right| \rho_{\rm ph}\left| U-\frac{X}{2} \right> \exp\left(-\frac{i}{2} X\Pi\right)\,{\rm d}X\ ,
	\end{equation}
which is a quasi-probability distribution in the phase space defined by the quadratures $\hat{u}$ and $\hat{\pi}$ and their respective eigenstates
	\begin{subequations}
\begin{align}
		&&\hat{u} &= \hat{b}+\hat{b}^\dagger\ ,&  \hat{\pi} &= \frac{1}{i} (\hat{b}-\hat{b}^\dagger)\ , &\\
		&&\hat{u}\left| U\right> &= U\left| U\right>\ ,&  \hat{\pi}\left| \Pi\right> &= \Pi\left| \Pi\right> \ .&
	\end{align}
	\end{subequations}

Due to their definition by the phonon annihilation and creation operators, the~quantities $U$ and $\Pi$ directly correspond to the lattice displacement and momentum, respectively. The~generating function~$F_\alpha$, at~the same time, is a characteristic function of the Husimi Q function~\cite{gerry2005intr}. From~this, we can directly calculate the instructive Wigner distribution analytically for a given pulse sequence via~\cite{schleich2011quan}:
\begin{equation}
		W(U,\Pi,t) = \frac{1}{4\pi^2} \iint\limits_{-\infty}^{\quad\infty} \exp\left(-\frac{|\alpha|^2}{2}\right) F_\alpha(t) \exp\left\{ i\left[ {\rm Re}(\alpha)\Pi +{\rm Im}(\alpha)U \right] \right\}\,{\rm d}^2\alpha\ . \label{eq:Wigner}
	\end{equation}

In the same way, we can isolate the Wigner function $W_{\rm x}$ for the phonons associated with the TLS being in the excited state $\left| x\right>$ by choosing $C_\alpha$ instead of $F_\alpha$ in Equation \eqref{eq:Wigner}. By~doing the same, but~choosing $Y_\alpha$, we define $W_{\rm p}$ as the Wigner function of the phonon assisted coherence. In~summary, we have
	\begin{subequations}\label{eq:Wigner_parts}
\begin{align}
		F_\alpha \to&\ W\ ,\\
		C_\alpha \to&\ W_{\rm x}\ ,\\ 
		Y_\alpha \to&\ W_{\rm p}\ ,\\
		F_\alpha-C_\alpha \to &\ W - W_{\rm x} =\ W_{\rm g}\ ,
	\end{align}
	\end{subequations}
where $W_{\rm g}$ is the Wigner function of the phonons associated with the TLS being in the ground state~$\left|g\right>\otimes\left|ph_{\rm g}\right>$.

Following the original definition by von Neumann~\cite{von1996math}, we investigate the time-dependent entropy of our coupled quantum system defined by
\begin{equation}
		S = -{\rm Tr}\big[ \rho \ln(\rho) \big]\ .
	\end{equation}

In general, subadditivity states that the entropy of the full system is a lower boundary for the sum of the entropies of the subsystems~\cite{wehrl1978gene}:
\begin{equation}
		S(\rho) \leq S(\rho_{\rm TLS} \otimes \rho_{\rm ph}) = S(\rho_{\rm TLS}) + S( \rho_{\rm ph})\ .
	\end{equation} 

It is important to note that for every pure quantum state $\rho_{\rm pure}$, the~entropy vanishes, i.e.,
\begin{equation}
		S_{\rm pure} = -{\rm Tr}\big[ \rho_{\rm pure} \ln(\rho_{\rm pure}) \big] = 0\ ,
	\end{equation}
and the entropies of the subsystems coincide if the state of the full system is pure~\cite{wehrl1978gene}
\begin{equation}
		S^{\rm ph} = S(\rho_{\rm ph} ) = S(\rho_{\rm TLS})\ . \label{eq:S_ph}
	\end{equation}

To show this, following Reference~\cite{wehrl1978gene} for an arbitrary composed system, we decompose the pure state of the entire system $\vert \psi \rangle$ into an orthonormal basis $\vert \psi \rangle = \sum_{i,k} C_{ik} \vert \phi_i \rangle \vert \chi_k \rangle$ with the coefficient matrix $C = (C_{ik})$. Here, $\vert \phi_i \rangle$ and $\vert \chi_k \rangle$ are basis states of the two subsystems, respectively. From~the complete density matrix $\rho = (\rho_{ikjl})$, one obtains a reduced density matrix by tracing over the respective other subsystem $\rho_1 = Tr_{2} (\rho)$: 
	\begin{subequations}
\begin{align}
		\rho_{ikjl} &= \langle \phi_i \vert \langle \chi_k \vert \rho \vert \chi_l \rangle \vert \phi_j \rangle =  C_{ik} C_{jl}^*\ ,\\ 
		\Rightarrow  \rho_{1,ij} &= \langle \phi_i \vert \rho_1 \vert \phi_j \rangle= \sum_k C_{ik} C_{jk}^* = (CC^\dagger)_{ij}\ .
	\end{align}
	\end{subequations}

Analogously the density matrix of the other subsystem is $\rho_2 = C^\dagger C$. Any non-vanishing eigenvalue $\lambda$ of $\rho_1$ with the eigenvector ${\bf y}$ is then also an eigenvalue of $\rho_2$ with eigenvector ${\bf z}=C^\dagger {\bf y}$ because
	\begin{subequations}\label{eq:eig}
\begin{align}
		CC^\dagger {\bf y} &= \lambda {\bf y}\ ,\\
		\Rightarrow C^\dagger C {\bf z} &= C^\dagger C(C^\dagger {\bf y}) = C^\dagger(CC^\dagger){\bf y} = \lambda C^\dagger {\bf y} = \lambda {\bf z}\ ,
	\end{align}
	\end{subequations}
and vice~versa.

In our particular system, we can choose the TLS's states as the $\vert \phi_i \rangle \in \{\vert g \rangle, \vert x \rangle\}$ and a Fock basis for the phonon system $\vert \chi_k\rangle \in \{ \vert 0 \rangle, \vert 1 \rangle,\ldots\}$. This means that the coefficient matrix $C$ consists of $2 \times \infty$ elements. The~entropy of the TLS, i.e.,~of its density matrix:  \vspace{6pt}
	\begin{subequations}
\begin{equation}
		CC^\dagger  = \rho_{\rm TLS} = \begin{pmatrix}
		1-c & p\\
		p^\ast & c
		\end{pmatrix}\ ,\qquad {\rm with}\quad c=\left<\left| x \right>\left< x\right| \right>\,,\ p=\left<\left| g \right>\left< x\right| \right>\ ,
	\end{equation}
can be easily calculated via~\cite{wehrl1978gene}
\begin{equation}\label{eq:S_TLS}
		S(\rho_{\rm TLS}) = -\lambda_+ {\rm ln}(\lambda_+)-\lambda_- {\rm ln}(\lambda_-)\ ,
	\end{equation}
where
\begin{equation}
		\lambda_{\pm} = \frac{1}{2} \pm \sqrt{\frac{1}{4}+c^2 - c +|p|^2} = \frac{1}{2} \pm \frac{1}{2} |{\bf v}|
		\label{eq:Bloch}
	\end{equation}
	\end{subequations}
are the two eigenvalues of the TLS's density matrix with the Bloch vector \mbox{${\bf v}=\big(2{\rm Re}(p),2{\rm Im}(p), 2c-1\big)$}. For~a pure state of the full system, according to Equation \eqref{eq:eig}, the~phonon density matrix has, despite~being a quadratic infinite dimensional matrix, only two non-vanishing eigenvalues $\lambda_\pm$. So,~as~long as the entire system is in a pure state and we know the entropy of the TLS via Equation~\eqref{eq:S_TLS}, we can derive the entropy of the phonon system by Equation~\eqref{eq:S_ph}.

For an arbitrary, non-pure state of the entire system, the~calculation of the entropy in a system with infinite dimensions is far from being trivial. This is the case if, already, the~initial state is a statistical mixture, e.g.,~at non-vanishing temperature or when dephasing leads to a statistical mixture. Therefore, approximations have been discussed and a reasonable version is given by the linear entropy~\cite{wehrl1978gene}:
\begin{equation}
		S_{\rm lin} = {\rm Tr}\big( \rho \big)  - {\rm Tr}\big( \rho^2 \big) = 1 - {\rm Tr}\big( \rho^2 \big) = 1-\left<\rho\right>\ .
		\label{eq:S_approx}
	\end{equation}

\begin{figure}[H]
	\centering
	\includegraphics[width=0.85\textwidth]{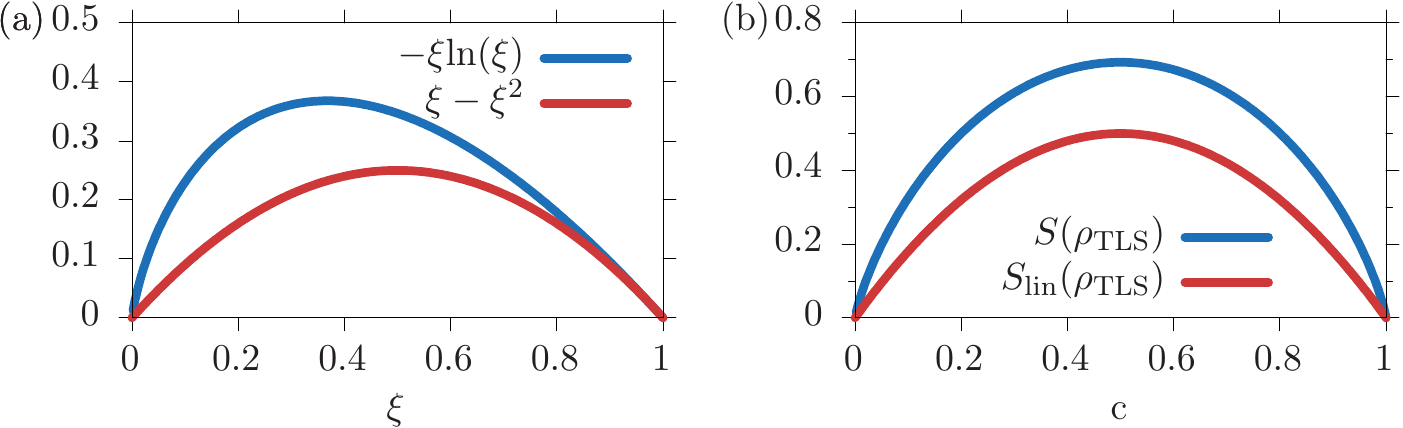}
	\caption{(\textbf{a}) Functions for the full entropy in blue and the linear entropy in red. (\textbf{b}) Entropies of the two-level system (TLS) with $p=0$, full entropy in blue and linear entropy in~red.}\label{fig:S_lin}
\end{figure}

To get an impression of this approximation, Figure~\ref{fig:S_lin}a shows the function of the full entropy $-\xi{\rm ln}(\xi)$ and the one for the linear entropy $\xi-\xi^2$; similar to the presentation in Reference~\cite{wehrl1978gene}, note that ${\rm Tr}(\rho)=1$. We find that the function of the approximated linear entropy in red is always smaller than the full entropy in blue. Therefore, we expect that the linear entropy under-estimates the full~entropy.
We can directly compare the linear and the full entropy of the isolated TLS from Equation \eqref{eq:S_TLS}, as~shown in Figure {\ref{fig:S_lin}b. There, the~full entropy is plotted in blue and the linear one in red as functions of the excited state occupation $c$, both for $p=0$. In~the limiting cases of full inversion $c=1$ and no inversion $c=0$, the~state is pure and both entropies vanish. For~all other occupations, the~TLS is in a statistical mixture, and~the entropy is non-zero, and~it is $S_{\rm lin}\leqslant S$. In~the case of an equally distributed mixture, i.e.,~$c=0.5$, the~entropies are maximal and reach values of $S_{\rm lin}= 0.5$ and $S=\ln(2)\approx 0.7$.

The biggest advantage of this linear entropy for our study is that it can be directly calculated from the phonons' Wigner functions due to the trace-product rule~\cite{manfredi2000entr} via
\begin{equation}
		S_{\rm lin}^{\rm ph} = 1 - 4\pi  \iint\limits_{-\infty}^{\quad \infty} W(U,\Pi)^2\,{\rm d}U\,{\rm d}\Pi \label{eq:S_lin_ph}\ .\\
	\end{equation}

Note that the prefactor $4\pi$ depends on the definition of the quadratures $U$ and $\Pi$, i.e.,~the scaling of the phase~space.

With the separation into TLS and phonon system, the~linear entropy is calculated via
\begin{equation}
		S_{\rm lin} = 1 -{\rm Tr}_{\rm ph}\left[ {\rm Tr}_{\rm TLS}\big( \rho^2 \big)\right]\ .
		\label{eq:S_lin_tr_tr}
	\end{equation}

For this, we again consider the density matrix of the full system as
	\begin{subequations}
\begin{align}
		\rho &= (1-c)\left| g \right>\left< g \right| \otimes \rho_{\rm ph,g} +  c\left| x \right>\left< x \right| \otimes \rho_{\rm ph,x} +  p^*\left| g \right>\left< x \right| \otimes \rho_{\rm ph,p} + p\left| x \right>\left< g \right|  \otimes \rho_{\rm ph,p}^\dagger\ ,\\
\Rightarrow {\rm Tr}_{\rm TLS}(\rho^2) &= (1-c)^2\rho_{\rm ph,g}^2 + c^2\rho_{\rm ph,x}^2 +\vert p \vert ^2 (\rho_{\rm ph,p}\rho_{\rm ph,p}^\dagger + \rho_{\rm ph,p}^\dagger \rho_{\rm ph,p})\ ,
	\end{align}
leading to
\begin{align}
		{\rm Tr}_{\rm TLS}(\rho) - {\rm Tr}_{\rm TLS}(\rho^2) &= (1-c) \rho_{\rm ph,g} - (1-c)^2 \rho_{\rm ph,g} \notag\\
									&+ c \rho_{\rm ph,x} - c^2 \rho_{\rm ph,x}^2 \notag\\
									&- |p|^2 (\rho_{\rm ph,p}\rho_{\rm ph,p}^\dagger + \rho_{\rm ph,p}^\dagger \rho_{\rm ph,p})\ .
	\end{align}

We can now use the separate parts of the Wigner function from Equation \eqref{eq:Wigner_parts} to define entropies
\begin{align}
		S_{\rm lin}^{\rm i} &= \iint\limits_{-\infty}^{\quad \infty}  \bigg[W_{\rm i}(U,\Pi) - 4\pi W_{\rm i}(U,\Pi)^2\bigg]\,{\rm d}U\,{\rm d}\Pi \ ,\qquad {\rm with}\ {\rm i}\in\{{\rm g,x }\}\ ,\\
		S_{\rm lin}^{\rm p} &= - 4\pi  \iint\limits_{-\infty}^{\quad \infty} \vert W_{\rm p}(U,\Pi)\vert^2\,{\rm d}U\,{\rm d}\Pi \ .
	\end{align}
	\end{subequations}

Note that the polarization Wigner function $W_{\rm p}$ is a complex quantity. With~this and the definitions of the generating functions in Equation \eqref{eq:gen_func}, we can write the linear entropy in Equation \eqref{eq:S_lin_tr_tr} as
\begin{equation}
		S_{\rm lin} = S_{\rm lin}^{\rm g} +S_{\rm lin}^{\rm x} + 2S_{\rm lin}^{\rm p} \ . \label{eq:S_lin}
	\end{equation}

To briefly summarize, for~pure states of the entire system, i.e.,~without dephasing or decay of the TLS, we can calculate the full entropy of the phonon state $S_{\rm ph}$ via Equation \eqref{eq:S_ph}. If~the state is not pure, we can at least calculate the linear entropy from the Wigner functions. We can further distinguish between the linear entropy of the full system $S_{\rm lin}$ in Equation \eqref{eq:S_lin} and the one of the phonons $S_{\rm lin}^{\rm ph}$ in Equation \eqref{eq:S_lin_ph}.

\section{Results and~Discussion}
\vspace{-6pt}
\subsection{Single Pulse~Excitation}
We start our study with the most basic situation, where the TLS is excited by a single optical pulse. It is well known from previous works that a single ultrafast excitation in general creates a statistical mixture of coherent states in the phonon system. The~excitation of the TLS means for the phonons a shift of the equilibrium position determined by the dimensionless coupling strength $\gamma=g/\omega_{\rm ph}$. If~not stated differently, in~the following, we fix this value to $\gamma=2$ in order to separate the different parts of the Wigner function in phase space, as~will be seen later. Although~$\gamma=2$ is a rather large value for quantum dots and optical phonons, the~general physics explained in this paper will not depend on this value. Some effects might be strengthened or weakened with a different choice of the coupling strength, as~will be highlighted later.
\subsubsection{Phonons Generated by a Non-Decaying~TLS}
In the first step, we neglect the decay of the excited state by choosing $\Gamma=0$. In~this situation, the~state of the full system, including the TLS and the phonons, is pure. Therefore, the~full and the linearized entropy are zero and Equation \eqref{eq:S_ph} holds, meaning that the entropy of the TLS and that of the phonons is the same. Figure~\ref{fig:1pulse} recapitulates the phonon dynamics for a pulse area of $\theta=\pi/2$, i.e.,~an~inversion of the TLS of $50\%$ or $c=0.5$, from~Reference~\cite{reiter2011gene}. The~phonon's Wigner function~reads:
\begin{align}
		W(U,\Pi,t) =& \frac{1}{4\pi}\Bigg\{ \exp\left[-\frac{1}{2}(U^2+ \Pi^2)\right] \\ 
		&+ \exp\left(-\frac{1}{2} \{ U-2\gamma [1-\cos(\omega_{\rm ph} t)] \}^2 -\frac{1}{2} [ \Pi-2\gamma \sin(\omega_{\rm ph} t) ]^2 \right) \Bigg\} \notag\ ,
			\label{eq:Wigner_coh}
	\end{align}
and its dynamics are shown at five different times in Figure~\ref{fig:1pulse}a. Before~the optical excitation, the~phonons are in the vacuum state represented by the Gaussian Wigner function in the center of the phase space. Half of the weight of the phonon's Wigner distribution is brought into the excited state subspace by the optical pulse. This makes them move as a coherent state around the new equilibrium position, which is shifted by $2\gamma$ in $U$-direction. This trajectory is marked as black circle in the figure. The~other half of the phonon state remains associated with the ground state of the TLS and, therefore, stays in the vacuum state. The~full phonon state after tracing over the TLS states is a statistical mixture of the vacuum state and a coherent state moving around the shifted equilibrium position. After~a full phonon period at $t=t_{\rm ph}$, the~Wigner function agrees with the initial situation because the coherent state moves through the origin and overlaps with the vacuum state. The~phonon's influence on the properties of the TLS is shown in Figure~\ref{fig:1pulse}b. While the occupation of the excited state stays constant at $c=0.5$, the~polarization $|p|$ starts at 0.5 directly after the optical excitation at $t=0$ and drops rapidly to almost zero in the following. This dephasing is inverted towards $t=t_{\rm ph}$, resulting in a full rephasing to $|p|=0.5$. While the coherent states separate in phase space, coherence~is lost from the TLS, which already shows that the overlap of the different parts of the Wigner function plays an important role for the properties of the entire~system.
\begin{figure}[H]
	\centering
		\includegraphics[width=\textwidth]{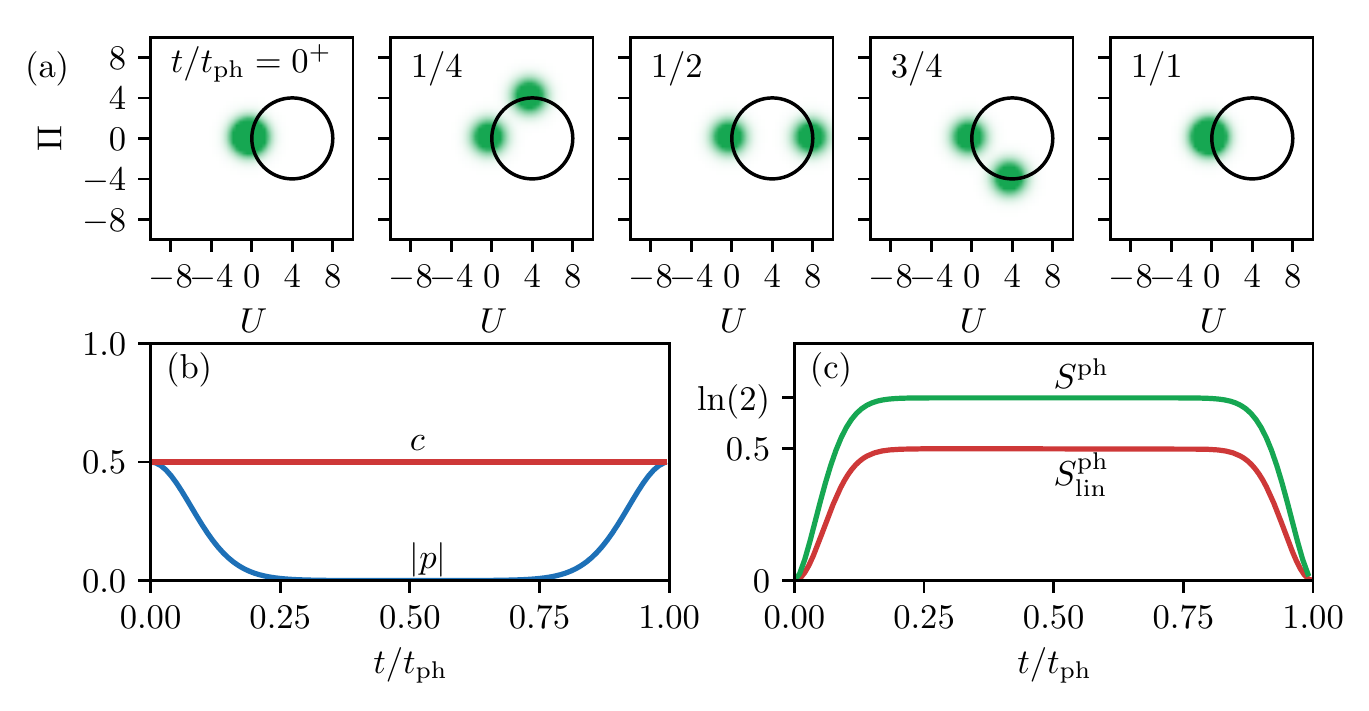}
	\caption{(\textbf{a}) Dynamics of the phonon Wigner function after a single pulse excitation of the TLS. (\textbf{b})~TLS dynamics with the excited state occupation $c$ in red and the polarization $|p|$ in blue. (\textbf{c}) Entropies of the phonon system, $S^{\rm ph}$ in green and $S^{\rm ph}_{\rm lin}$ in~red.}\label{fig:1pulse}
\end{figure}
Finally, in~Figure~\ref{fig:1pulse}c, we show the entropy of the phonons which, as~mentioned above, agrees with the entropy of the TLS. As~the initial phonon state is pure, both entropies, $S^{\rm ph}$ in green and $S^{\rm ph}_{\rm lin}$ in red, start at zero at $t=0$. While the different parts in phase space separate, the~entropy grows to $S^{\rm ph}_{\rm lin}\approx 0.5$ because the phonons are in a statistical mixture that must have a non-vanishing entropy. The~full entropy follows the same dynamics as the linear one but is always larger, as~previously explained, and~grows to $S^{\rm ph}\approx\ln(2)\approx 0.7$. Reaching $t=t_{\rm ph}$, the~entropies drop to zero again. The~reason is the recovered overlap of the two parts of the Wigner function. Finally, at~$t=t_{\rm ph}$, the~phase space representation cannot be distinguished from the vacuum state. Therefore, the~entropy also has to agree with the one of the pure vacuum state being~zero.

\subsubsection{Phonons Generated by a Decaying~TLS}
In the previous section, without~any decay or additional pure dephasing of the TLS, the~quantum state of the entire system remained pure, resulting in a vanishing entropy of the full system. It also allowed us to easily calculate the full entropies for TLS and phonons. In~this section, we consider a non-vanishing decay rate of the occupation of the excited state into the ground state, which naturally results in a statistical mixture in the TLS's quantum state that also imprints onto the phonons. Therefore, for~the phonons, we can only calculate linear entropies, according to Equation \eqref{eq:S_lin_ph}. In~Reference~\cite{hahn2019infl}, we explained how the Wigner function evolves during the decay process in the TLS. Therefore, we~consider the same optical excitation with a pulse area of $\theta=\pi$, which initially fully inverts the TLS. Without~any decay, the~Wigner function would read
\begin{equation}
		W(U,\Pi,t) = \frac{1}{2\pi}\exp\left(-\frac{1}{2} \{ U-2\gamma [1-\cos(\omega_{\rm ph} t)] \}^2 -\frac{1}{2} [ \Pi-2\gamma \sin(\omega_{\rm ph} t) ]^2 \right)\ , \label{eq:Wigner_coh}
	\end{equation}
being a single Gaussian moving on a circle around the shifted equilibrium position of the excited state. In~Figure~\ref{fig:1pulse_decay}a, Wigner functions for different decay rates $\Gamma$ are shown at $t=10t_{\rm ph}$. We find that the phonon state gets smeared out in phase space. For~rapid decays on the left, the~phonons almost completely stay in the vacuum state and look more or less like a coherent Gaussian distribution. When looking at the corresponding linear entropy dynamics in Figure~\ref{fig:1pulse_decay}b, in~bright red, we see that it only increases slightly after the optical excitation at $t=0$. When slowing down the decay process, i.e.,~moving in Figure~\ref{fig:1pulse_decay}a more to the right, the~Wigner function smears out more and more. Accordingly it looks less and less like a coherent state which also leads to increasing entropies in (b) when going from bright to dark colors. Additionally, we find that the final entropy value is reached slower because it follows the decay of the TLS. Especially for the slowest considered decay of $\Gamma=0.1\omega_{\rm ph}$, where the full decay takes several phonon periods, the~dynamics of the linear phonon entropy develop minima at full phonon periods $t=n t_{\rm ph}$. These are the times when the Wigner function in Figure~\ref{fig:1pulse_decay}a starts overlapping itself. This is exemplarily shown in Figure~\ref{fig:1pulse_decay}c for $t=t_{\rm ph}$, where the thick Gaussian part is the oscillating coherent state. This is the first time it intersects with the circular distribution that has already decayed into $\left| g\right>$. In~agreement with the findings in Figure~\ref{fig:1pulse}, this leads to the temporary reduction of the entropy in Figure~\ref{fig:1pulse_decay}b.
\begin{figure}[H]
\centering
\includegraphics[width=\textwidth]{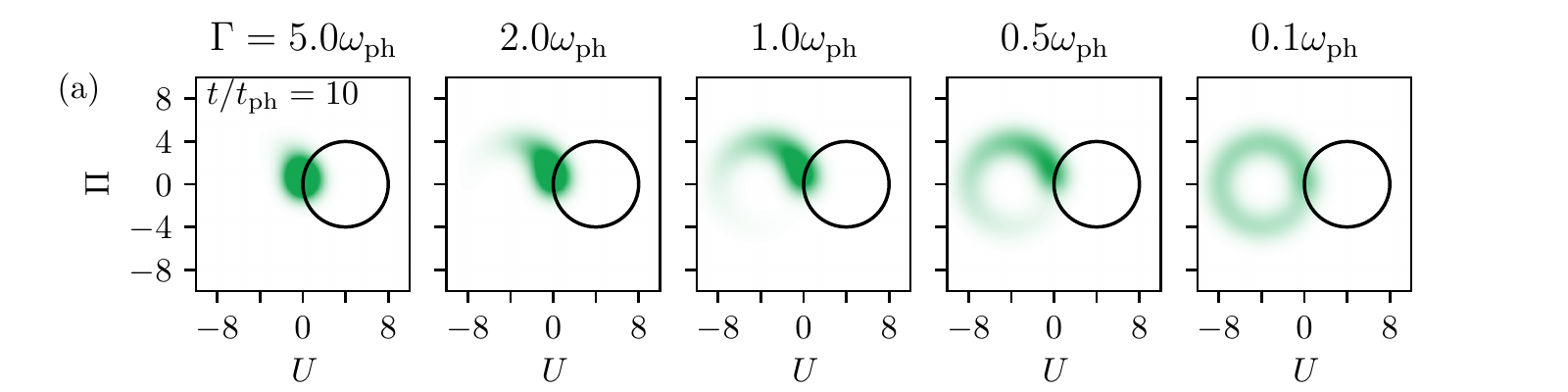}\\
\includegraphics[width=\textwidth]{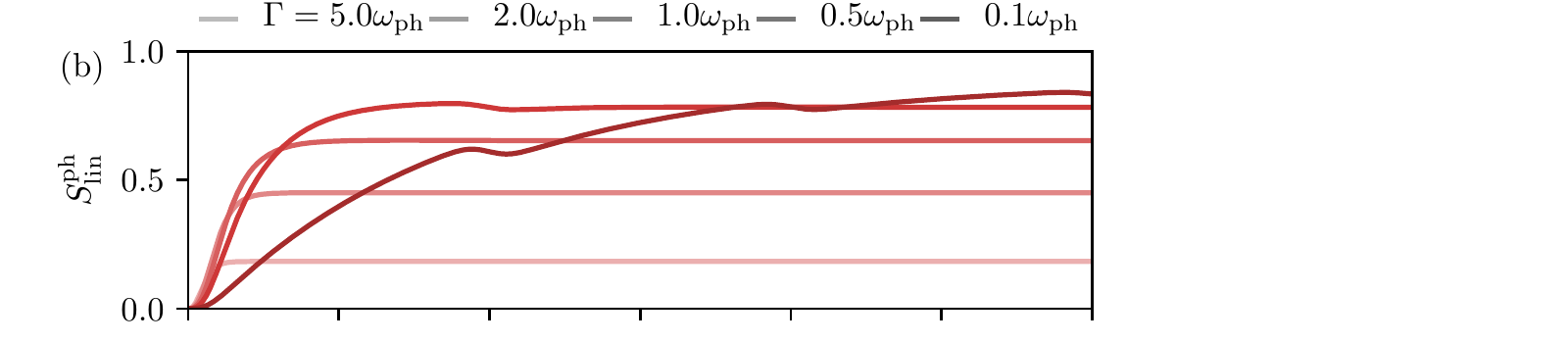}
\hspace{-.28\textwidth}\includegraphics[width=.24\textwidth]{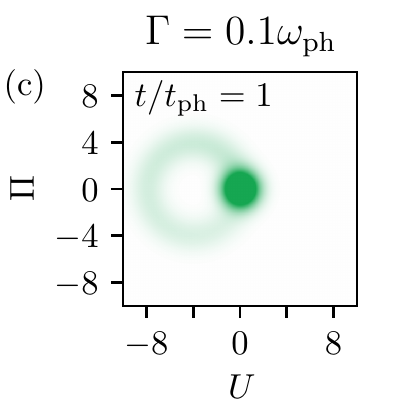}\\
~\hspace{0.013\textwidth}\includegraphics[width=\textwidth]{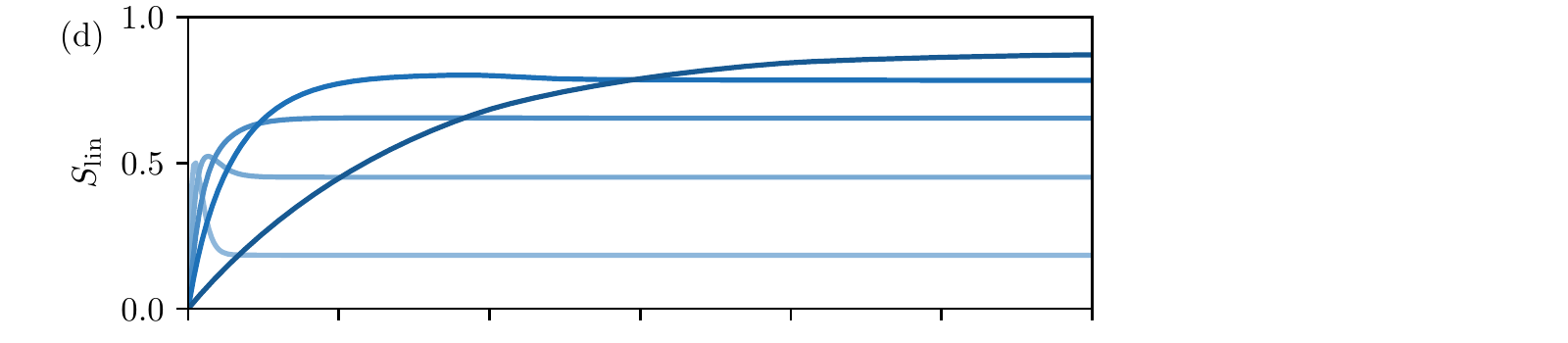}\\[-2mm]
\hspace{-.02\textwidth}\includegraphics[width=\textwidth]{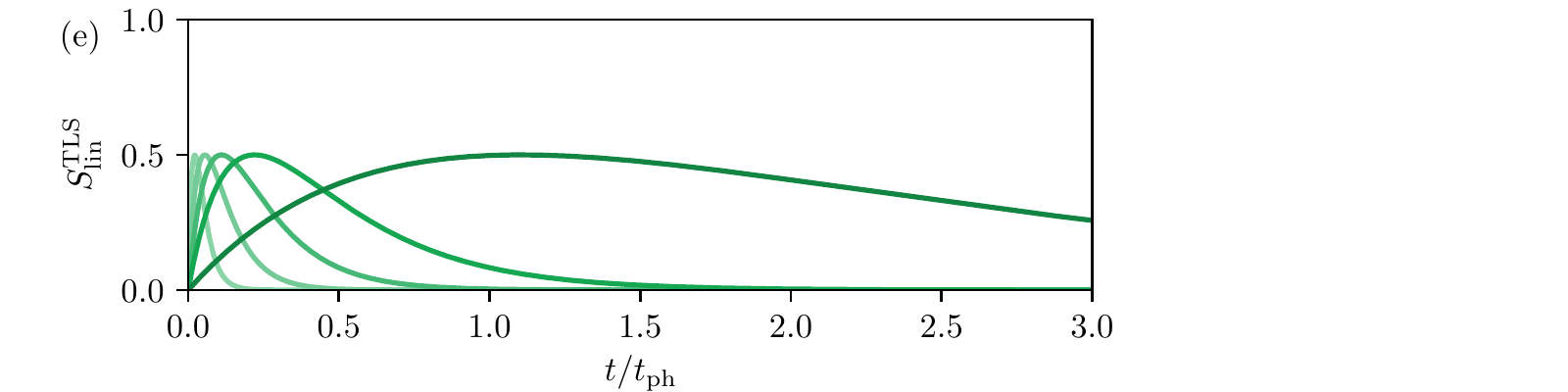}
\hspace{-0.28\textwidth}\includegraphics[width = .22\textwidth]{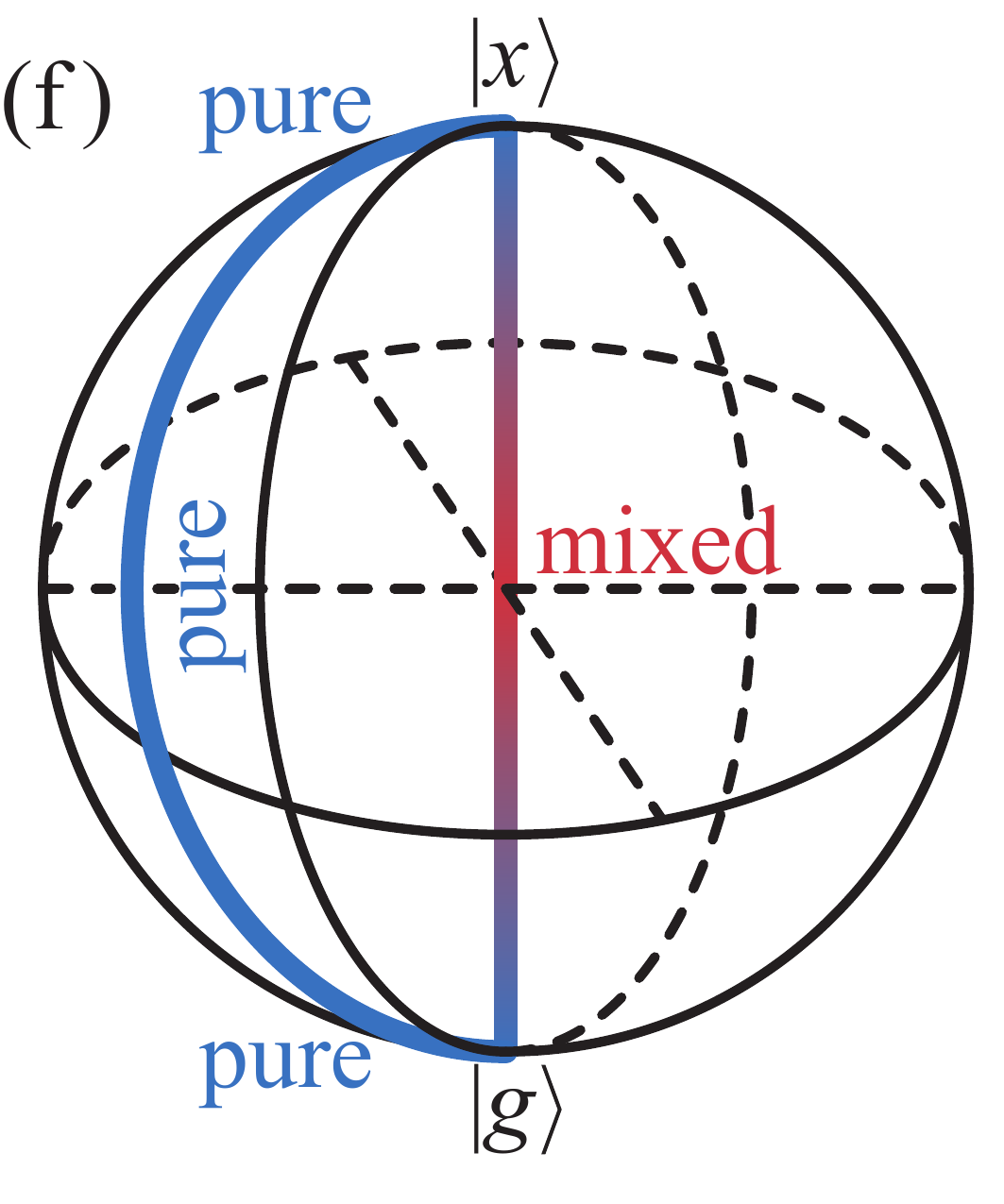}
\caption{(\textbf{a}) Phonon Wigner function after a full decay of the TLS at $t=10t_{\rm ph}$ for different decay rates $\Gamma$ as given in the picture. (\textbf{b}) Linear phonon entropy as a function of time after the pulse. The~decay rate decreases from bright to dark colors. (\textbf{c}) Exemplary Wigner function for a slow decay rate of $\Gamma=0.1\omega_{\rm ph}$ at $t=t_{\rm ph}$. (\textbf{d}) Same as (\textbf{b}) but for the full linear entropy. (\textbf{e}) Same as (\textbf{b}) but for the linear entropy of the TLS. (\textbf{f}) Bloch sphere of the TLS to illustrate the purity of different states; blue shows pure and red mixed~states.}\label{fig:1pulse_decay}
\end{figure}
Comparing the linear entropy of the phonon system in Figure~\ref{fig:1pulse_decay}b with the linear entropy of the entire system $S_{\rm lin}$ in (d), we basically find the same overall behavior. The~dynamics start at zero, the~final values increase for smaller decay constants, and~the final values are reached later. However, both for small and large $\Gamma$, we find qualitative differences. Starting with small $\Gamma$ in dark blue, especially for $\Gamma=0.1\omega_{\rm ph}$, the~curve constantly grows without developing any minima. This shows that the reduction of phonon entropy due to the overlapping Wigner functions does not attain to the full entropy. The~reason for this is that the reduction of the phonon entropy due to overlapping parts of the Wigner function only happens because the state information of the TLS has been traced out. Taking the entire coupled system into account, the~overlapping phonon parts belong to different states of the TLS and can therefore be told apart. Therefore, the~phonons do not lead to a depression of the full linear entropy. Conversely, for~large decay rates, in~bright blue, especially $\Gamma=5\omega_{\rm ph}$, we find that, on~a short timescale around $t=0.1 t_{\rm ph}$, a~pronounced maximum appears in $S_{\rm lin}$. This effect is not found in the phonon part in Figure~\ref{fig:1pulse_decay}b and therefore stems from the TLS contribution. To~understand this, in~Figure~\ref{fig:1pulse_decay}e, we plot the linear entropy of the TLS as green lines. The~bright and dark colors agree with the ones in Figure~\ref{fig:1pulse_decay}b,d. In~addition, we consider the schematic Bloch vector representation of the TLS state from Equation \eqref{eq:Bloch} in Figure~\ref{fig:1pulse_decay}f. The~$z$ direction of the Bloch sphere depicts the occupation of the states, where the south pole is a pure ground state $\left| g\right>$ and the north pole a pure excited state $\left| x\right>$. Between~these points, all Bloch vectors that are on the surface of the sphere (blue line) are superpositions $\left| \chi\right> = N (\alpha\left|g\right> + \beta \left|x\right>)$ and are therefore pure. In~the other extreme case of the line directly connecting north and south pole (red), the~system is in a statistical mixture of $\left| g\right>$ and $\left| x\right>$. In~the center of the Bloch sphere, the~TLS is in both states with equal probability, resulting in the lowest purity. For~the entropy of the TLS in Figure~\ref{fig:1pulse_decay}e, this means that, directly after the excitation into the excited state and after the full decay, the~entropy is zero. In~between, the~system evolved through a statistical mixture, which has a non-vanishing entropy. The~linear entropy reaches maxima with $S_{\rm lin}^{\rm TLS}= 0.5$, in~agreement with the result in Figure~\ref{fig:S_lin}b.

If we want to determine the final linear entropy $S_{\rm lin}^{\rm ph, \infty}=S_{\rm lin}^{\rm ph}(t\to\infty)$ of the phonon state after the TLS is fully decayed into the ground state, we can investigate the dynamics in phase space. Note~that the final entropy is only carried by the phonon part because the TLS is in the pure ground state. As~schematically shown in Figure~\ref{fig:final_S}a, the~movement of the coherent state on the circle in the excited state subspace and the accompanied decay into the ground state leads to a distribution that can be seen as a continuous distribution of coherent states with decreasing amplitude. We can parametrize the circular motion of the Wigner function including the decay of the amplitude by
\begin{equation}
		W_{\infty}(U,\Pi) = \Gamma \int\limits_0^\infty \exp(-\Gamma t) W_{(U_0(t),\Pi_0(t))} \,{\rm d}t\ ,
	\end{equation}
where $W_{(U_0(t),\Pi_0(t))}$ is a Gaussian centered around $(U,\Pi)=(U_0(t),\Pi_0(t))$. With~the circular trajectory in Equation \eqref{eq:Wigner_coh}, we have to consider
\begin{equation}
		U_0(t) = 2\gamma [1-\cos(\omega_{\rm ph}t)] \qquad {\rm and}\qquad
		\Pi_0(t) = 2\gamma \sin(\omega_{\rm ph}t) \ . \label{eq:U0_Pi0}
	\end{equation}

Note that, to~retrieve the Wigner distribution in the ground state in Figure~\ref{fig:1pulse_decay}a, one has to mirror the schematic in Figure~\ref{fig:final_S}a. However, the~final linear entropy remains unaffected because it only depends on the general shape of the distribution.
With this, the~final linear entropy reads
\begin{align}
		S_{\rm lin}^{\rm ph, \infty}
		&= 1-4\pi\iint \limits_{-\infty}^{\quad\infty} W_{\infty}^2(U,\Pi)\, {\rm d}U{\rm d}\Pi \notag\\
		&= 1-4\pi \Gamma^2 \iint \limits_{-\infty}^{\quad\infty} \iint \limits_{0}^{\quad\infty} \exp\left[-\Gamma (t+t^\prime)\right]  W_{(U_0(t),\Pi_0(t))}   W_{(U_0(t^\prime),\Pi_0(t^\prime))}\, {\rm d}t {\rm d}t^\prime  \, {\rm d}U{\rm d}\Pi \ .
	\end{align}

The integral over $U$ and $\Pi$ describes the overlap of two coherent states in phase space. In~general, two coherent states with a phase space distance of $a$ have an overlap of~\cite{schleich2011quan}
\begin{equation}
		4\pi\iint \limits_{-\infty}^{\quad\infty} W_{(0,0)}W_{(a,0)}(U,\Pi)\, {\rm d}U{\rm d}\Pi = \exp\left[-\frac{a^2}{4}\right]\ .
	\end{equation}

Therefore, the~entropy becomes
\begin{equation}
		S_{\rm lin}^{\rm ph, \infty} = 
			1 - \Gamma^2  \iint \limits_{0}^{\quad\infty} \exp\left[-\Gamma (t+t^\prime)\right] \exp\left(-\left\{2\gamma\sin\left[\frac{1}{2}\omega_{\rm ph}(t-t^\prime)\right]\right\}^2\right)\, {\rm d}t {\rm d}t^\prime\ . \label{eq:S_lin_inf}
	\end{equation}

The sine function in the exponent can be approximated by a linear function for small frequencies or short times. Note that, although~the integration is carried out up to $t=\infty$, the~exponential decay with $\Gamma$ effectively limits the integrated time interval. Therefore, we expect this approximation to work well for sufficiently large $\Gamma$. For~the motion of the Wigner function, this corresponds to a linear movement in phase space, as~schematically shown in Figure~\ref{fig:final_S}b. The~corresponding linear entropy can then be calculated to 
\begin{align}
		S_{\rm lin}^{\rm ph, \infty}
		&\approx  1 - \Gamma^2  \iint \limits_{0}^{\quad\infty} \exp\left[-\Gamma (t+t^\prime)\right] \exp\left\{-[\gamma\omega_{\rm ph}(t-t^\prime)]^2\right\}\, {\rm d}t {\rm d}t^\prime \notag\\
		&= 1 - \frac{\sqrt{\pi}\,\Gamma}{2\gamma\omega_{\rm ph}} \exp\left[\left(\frac{\Gamma}{2\gamma\omega_{\rm ph}}\right)^2\right] {\rm erfc}\left( \frac{\Gamma}{2\gamma\omega_{\rm ph}} \right)\ ,
		\label{eq:S_inf_approx}
	\end{align}
where ${\rm erfc}(x) = 1-{\rm erf}(x)$ and ${\rm erf}(x)$ are the error~function.
\begin{figure}[H]
	\centering
		\includegraphics[width=\textwidth]{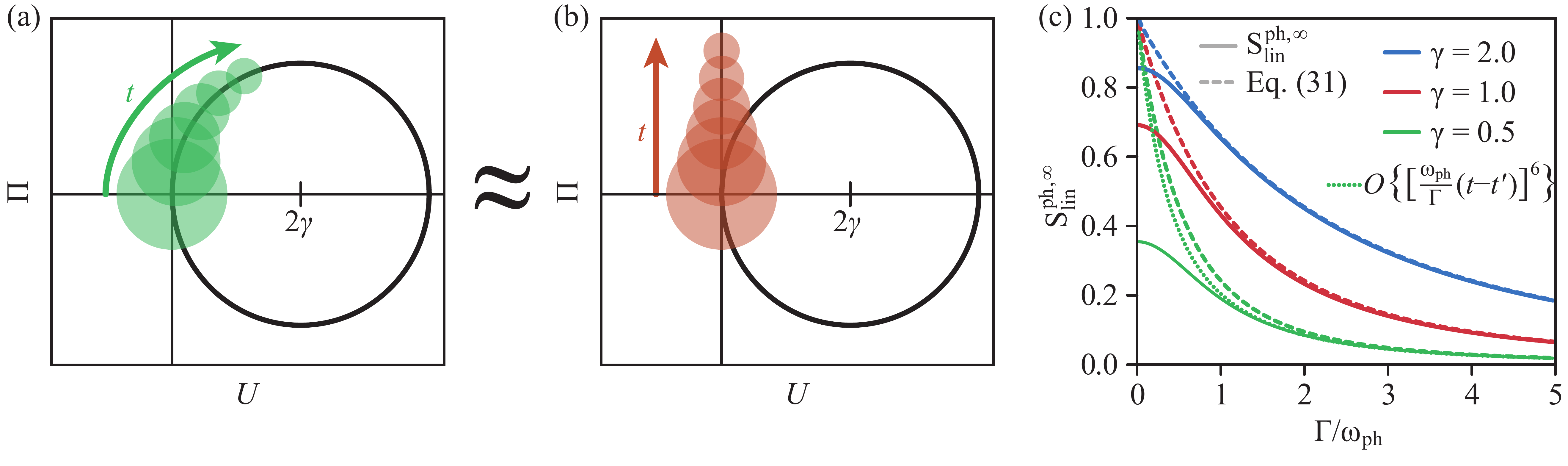}
	\caption{Final value of the entropy after the full decay of the TLS. (\textbf{a}) Schematic of the phase space dynamics of the Wigner function during the decay. (\textbf{b}) Approximated dynamics as a straight line. (\textbf{c})~Final linear entropy $S_{\rm lin}^{\rm ph,\infty}$ as a function of the decay rate $\Gamma$, full simulation in solid and approximation from Equation \eqref{eq:S_inf_approx} in dashed lines. Different coupling strengths are shown in blue, red, and~green. The~dotted green line shows the entropy for an expansion of the squared sine-function up to the sixth~order.}
	\label{fig:final_S}
\end{figure}

The results for the final entropy $S_{\rm lin}^{\rm ph,\infty}$ are shown in Figure~\ref{fig:final_S}c as a function of the decay rate $\Gamma$, where the full calculations according to the dynamics are shown as solid lines and the approximations from Equation \eqref{eq:S_inf_approx} are the dashed lines. We show the three different coupling strengths $\gamma=2$ (blue), $\gamma=1$ (red), and~$\gamma=0.5$ (green). Comparing the different coupling strengths, we find that a stronger coupling leads to a larger final entropy because the Wigner function gets distributed over a larger area of phase space. As~explained before, the~approximation in Equation \eqref{eq:S_inf_approx} works very good for large $\Gamma$, but~we also see that the approximation works over a larger $\Gamma$ range if the coupling strength is larger. The~reason for this is that, for~larger $\gamma$, the~circle of the Wigner function's trajectory is larger, meaning~that its curvature can be better approximated by a linear motion. For~the smallest considered coupling strength $\gamma=0.5$, in~green, we additionally show the dotted line that stems from an approximation of the squared sine function up to the sixth order, which is the next non-divergent contribution in Equation \eqref{eq:S_lin_inf}. This curve is obviously a better approximation of the full calculation, in~particular, in~the range $1\lesssim \Gamma/\omega_{\rm ph}\lesssim 2$. However, in~agreement with all dashed lines, it reaches $S_{\rm lin}^{\rm ph,\infty}=1$ for $\Gamma=0$, while all full linear entropies go to smaller values. This shows that any approximation of the sine function will not give accurate results for very small decay rates. In~fact, it turns out that, for~any finite (converging) order of the expansion of the squared sine function in Equation \eqref{eq:S_lin_inf}, $S_{\rm ph}^{\rm lin,\infty} = 1$ is reached in the limit $\Gamma \to 0$. This can be understood by realizing that, for~any finite order in the expansion of the trigonometric functions in Equation \eqref{eq:U0_Pi0}, the~trajectory tends to infinity for $t \to \infty$, thus leading to a delocalized Wigner function. On~the other hand, the~correct Wigner function in this limit is the doughnut-shaped function similar to the rightmost function in Figure~\ref{fig:1pulse_decay}a, for~which the linear entropy can be calculated analytically, yielding
\begin{equation}
		S_{\rm ph}^{\rm lin,\infty} = 1 - e^{-2\gamma^2} I_0\big (2\gamma^2\big )\ ,
	\end{equation}
with the modified Bessel function of first kind and zeroth order $I_0$, in~perfect agreement with the numerical results given in Figure~\ref{fig:final_S}c.

\subsection{Two Pulse~Excitation}
The phonon quantum state gets more involved when a two-pulse excitation is considered. As~extensively studied in Reference~\cite{reiter2011gene,hahn2019infl}, an~excitation with two pulses having pulse areas of $\theta_1=\theta_2=\pi/2$ and a delay of $t_2-t_1=t_{\rm ph}/2$ leads to the generation of two Schrödinger cat states, each in one TLS subspace. A~Schrödinger cat state is a coherent superposition of two coherent states, i.e.,~of the most classical states of a harmonic oscillator, and, as~such, it is of high interest in all areas of quantum optics~\cite{brune1992mani,ourjoumtsev2007gene,deleglise2008reco} and, more recently, also phononics~\cite{hofheinz2009synt}.

\subsubsection{Phonons Generated by a Non-Decaying~TLS}\label{sec:2pulse_no_decay}
We assume the same excitation scheme as just described and again disregard any decays of the TLS. The~dynamics of the Wigner function are exemplarily shown in Figure~\ref{fig:2pulse}a. Immediately before the second laser pulse reaches the TLS, the~phonons are in the statistical mixture previously shown in Figure~\ref{fig:1pulse}a. The~second pulse creates a second coherent state in the excited state subspace of the TLS, but~it also makes half of the coherent state in $\left| x\right>$ go back to the ground state. Therefore, we end up with two coherent states in both subspaces, $\left| g\right>$ and $\left| x\right>$. As~nicely seen in Figure~\ref{fig:2pulse}a, the~corresponding Wigner function shows two of the classic dumbbell structures of the cat state, two~Gaussians and a striped structure of alternating positive (green) and negative (orange) values between them. These~stripes indicate the interference between the two coherent states. The~Wigner function in the ground state rotates around the origin, and~the one in the excited state around the shifted equilibrium at $(U,\Pi)=(2\gamma,0)=(4,0)$ is marked by the black~circle. \vspace{-12pt}
\begin{figure}[H]
	\centering
		\includegraphics[width=\textwidth]{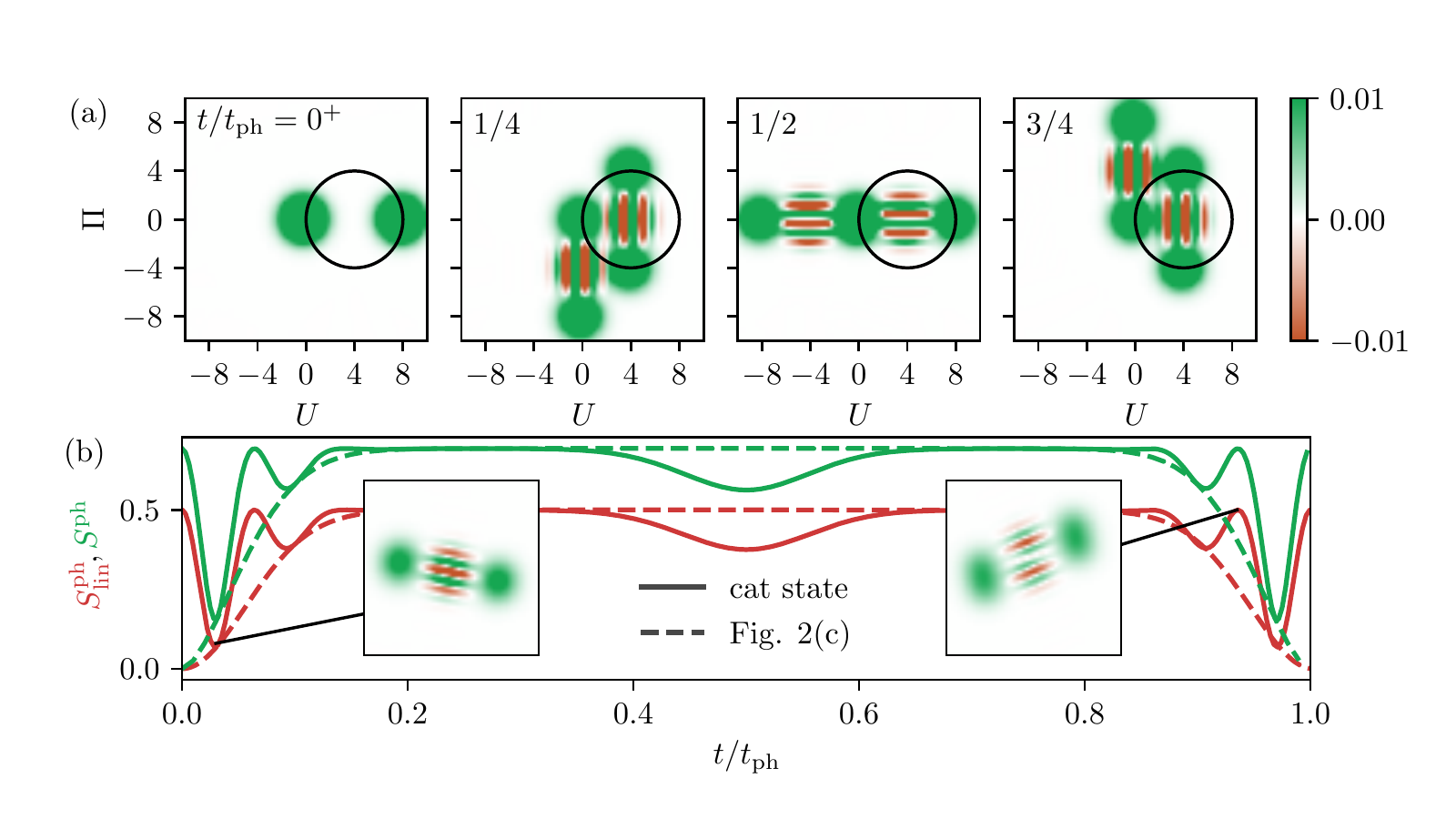}
	\caption{Two pulse excitation without decay. (\textbf{a}) Snapshots of the Wigner function after the second pulse. (\textbf{b}) Entropy dynamics after the second pulse, solid lines are the entropies of the cat states, dashed~lines the statistical mixture from Figure~\ref{fig:1pulse}c. The~full phonon entropy is green and the linear one~red.}\label{fig:2pulse}
	\end{figure}

The entropies of the phonon state are shown in Figure~\ref{fig:2pulse}b as solid lines. The~linear entropy in red starts at $S_{\rm lin}=0.5$ in agreement with the entropy in Figure~\ref{fig:1pulse}d at $t=t_{\rm ph}/2$ because at this time the second pulse excites the TLS. In~contrast to the behavior after the first pulse, here, the~entropy drops very rapidly and forms a sharp minimum after the second pulse. In~total, the~entropy performs two oscillations before remaining constant at the initial value. The~same dynamics repeat themselves in an inverted form before reaching a full period. Another striking feature is a smaller depression around $t=t_{\rm ph}/2$. Overall, we find that, by~the second pulse, the~entropy of the phonon is temporarily reduced but never increased. The~maximum entropy is here the one of a statistical mixture of two fully separated coherent states (see Figure~\ref{fig:1pulse}), which is obviously the same for a statistical mixture of two cat states that are fully separated. This is the case during the times around $t/t_{\rm ph}=1/4$ and $t/t_{\rm ph}=3/4$ (see Figure~\ref{fig:2pulse}a). Next, we discuss the reduced entropy around half a period in Figure~\ref{fig:2pulse}b. Looking at the corresponding Wigner function in Figure~\ref{fig:2pulse}a, we find that this is the time when one of the coherent states in the excited state system (moving on the circle) overlaps with the vacuum state (staying at the origin). This is the same effect as discussed in the previous sections, where the entropy shrank when the phonon states were overlapping in phase space. \mbox{Finally, we have} to understand the strong reductions of the entropy for times around full periods. To~do so, we examine the two insets in Figure~\ref{fig:2pulse}b that show snapshots of the Wigner function at the marked times, i.e.,~where~the entropy is minimal and maximal. The~left one at the minimum depicts a time where each of the two Gaussians starts to split into two, which cannot yet be resolved in the figure because their overlap is still too large. However, the~interference terms between them also move apart. At~$t=0$, their~negative and positive values are distributed in such a way that they exactly compensate each other (see $t=0^+$ in Figure~\ref{fig:2pulse}a). But, in~the left inset, we see that at this time negative and positive values add up, respectively, making~for an accurate alternating pattern. Comparing this structure of the Wigner function with one of the cat states in Figure~\ref{fig:2pulse}a shows a strong resemblance. So, the~reason for the strong decrease of the entropy is that, at~these times, the~Wigner function can only hardly be distinguished from a single Schrödinger cat state, which is a pure state. Likewise, we can analyze the Wigner function at a maximum of the entropy oscillation in the right inset. In~addition, here, the~Gaussians have a large overlap, but~the stripes of the interferences are aligned in such a way that the line in the center has vanishing values. This strongly disagrees with the natural structure of a cat state interference and makes it easily distinguishable from that pure cat state. The~full phonon entropy $S^{\rm ph}$ is shown as a green solid line. It follows the same dynamics as the linear one but is just scaled to larger values, as~discussed before. Finally, let us remark on the additionally plotted dashed lines, which are the respective entropy curves from Figure~\ref{fig:1pulse}c. They~exactly form envelopes for the oscillations and therefore demonstrate that the oscillating dynamics are again a result of the separation process of the different Wigner functions in phase~space.

The relative phase of the two laser pulses changes the phase in the cat states, i.e.,~the phase of the striped structure of the Wigner function. As~long as the different parts of the phonon state are separated in phase space, the~phase has no influence on the phonon entropy. The~other crucial parameter of the phonon system is the coupling strength $\gamma$ that determines the distance of the coherent states and the number of stripes in the interference term, as~exemplarily shown in Figure~\ref{fig:S_gamma}a,b. The~influence of an increased coupling strength is presented in Figure~\ref{fig:S_gamma}c, where the linear phonon entropy $S^{\rm ph}_{\rm lin}$ is plotted in the same way as in Figure~\ref{fig:2pulse}b but only for times up to $t/t_{\rm ph}=0.25$. The~red curve shows $\gamma=2$ from Figure~\ref{fig:2pulse}b as a reference. Looking at the larger coupling in bright red ($\gamma=4$) and a smaller coupling in dark red ($\gamma=1.5$), we clearly see that the oscillation of the entropy gets faster when $\gamma$ grows and more minima evolve. The~reason is that the interference terms consist of more stripes that run through each other. At~the same time, the~envelope gets shorter for a larger $\gamma$, which can be traced back to the larger spread of the Wigner function in phase space. While the sizes of the interference terms and the Gaussians stay the same, as~shown in Figure~\ref{fig:S_gamma}a,b, the~radius of the trajectory increases. Because~the angular frequency of the motion remains the same, the~two interferences separate in a shorter time, and~this time determines the envelope of the entropy in Figure~\ref{fig:S_gamma}c.
\begin{figure}[H]
	\centering
		\includegraphics[width=\textwidth]{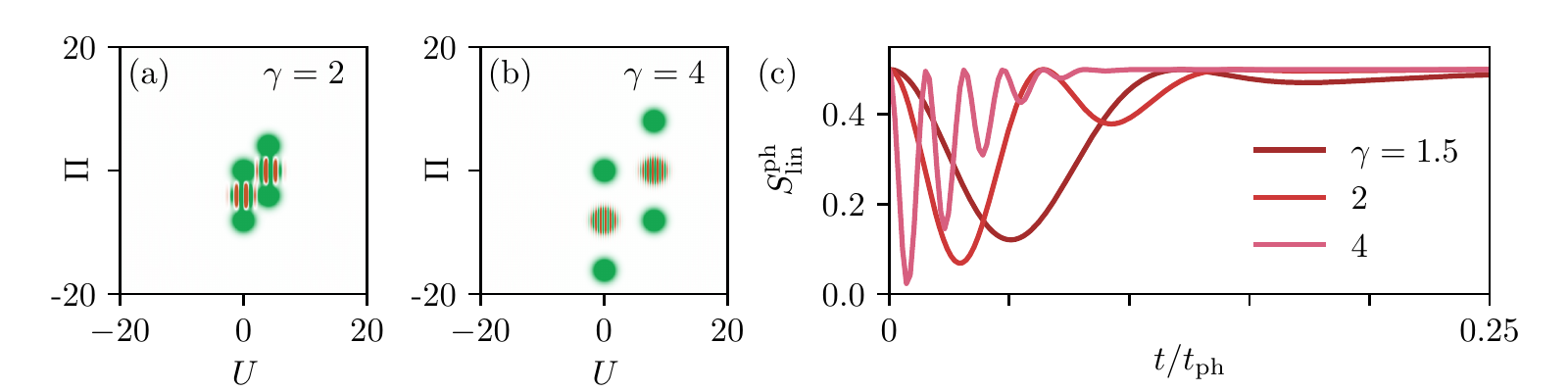}
	\caption{(\textbf{a},\textbf{b}) Exemplary Wigner functions at $t=t_{\rm ph}/4$ for $\gamma=2$ in (\textbf{a}) (same as in Figure~\ref{fig:2pulse}a) and $\gamma=4$ in (\textbf{b}). (\textbf{c}) Linear phonon entropy $S^{\rm ph}_{\rm lin}$ as a function of time after a two-pulse excitation as in Figure~\ref{fig:2pulse}b. The~coupling strength $\gamma$ increases from dark to bright~red.}\label{fig:S_gamma}
\end{figure}
\unskip

\subsection{Phonon Cat State Entropy Dynamics in a Decaying~TLS}
Next, we increase the complexity of the considered phonon state by analyzing the influence of the decay of the TLS on the entropy dynamics of cat states. For~reasons of clarity, we consider a single Schr\"odinger cat state entirely in the excited state $\left| x\right>\otimes\left|{\rm cat}\right>$ as initial state without any optical excitation and account for a decay of the TLS into its ground state. Although~this state cannot directly be prepared by optical pulses, in~Reference~\cite{hahn2019infl} it is explained how it is constructed mathematically as initial state for the simulated decay dynamics. Some snapshots of the corresponding Wigner function dynamics are shown in Figure~\ref{fig:decay_cat}a for a small decay rate of $\Gamma=0.1\omega_{\rm ph}$. As~analyzed in Reference~\cite{hahn2019infl}, the~combined rotation and shift of the phonon equilibrium position due to the decay of the TLS finally leads to a Wigner function in the shape of an eight. Note that, for~a slow decay where the coherent parts lead to a homogeneously distributed eight-shape, two interference terms transferred into the ground state at $t/t_{\rm ph} =(2n+1)/4$ survive the decay~process.

The corresponding linear entropies are depicted in Figure~\ref{fig:decay_cat}b, where the blue curves show the linear entropy of the full system $S_{\rm lin}$ and the red ones the linear entropy of the phonons $S_{\rm lin}^{\rm ph}$. The~decay rate increases from dark to bright colors. We find the same dependency on the decay rate as for a single coherent state in Figure~\ref{fig:1pulse_decay}b,d, the~final entropy increases for a slower decay. In~addition, the~behavior for very fast decays, e.g.,~$\Gamma=5\omega_{\rm ph}$, is approximately the same as in Figure~\ref{fig:1pulse_decay}. The~entropy of the full system (light blue) forms a sharp peak due to the evolution of the TLS through a statistical mixture, while the phonon part (light red) basically just rises before reaching the stationary value. The~dynamics get more involved and new features appear for slow decays, e.g.,~for $\Gamma=0.05\omega_{\rm ph}$. Here, the~full entropy in dark blue continuously increases to the stationary value at the end of the decay process, while the phonon contribution in dark red is always slightly smaller and shows additional dynamics developing multiple minima and maxima within each phonon period. While the dynamics are rather irregular on shorter times $t< 3t_{\rm ph}$, it becomes more periodic for longer periods of~time.
\begin{figure}[H]
	\centering
		\includegraphics[width=\textwidth]{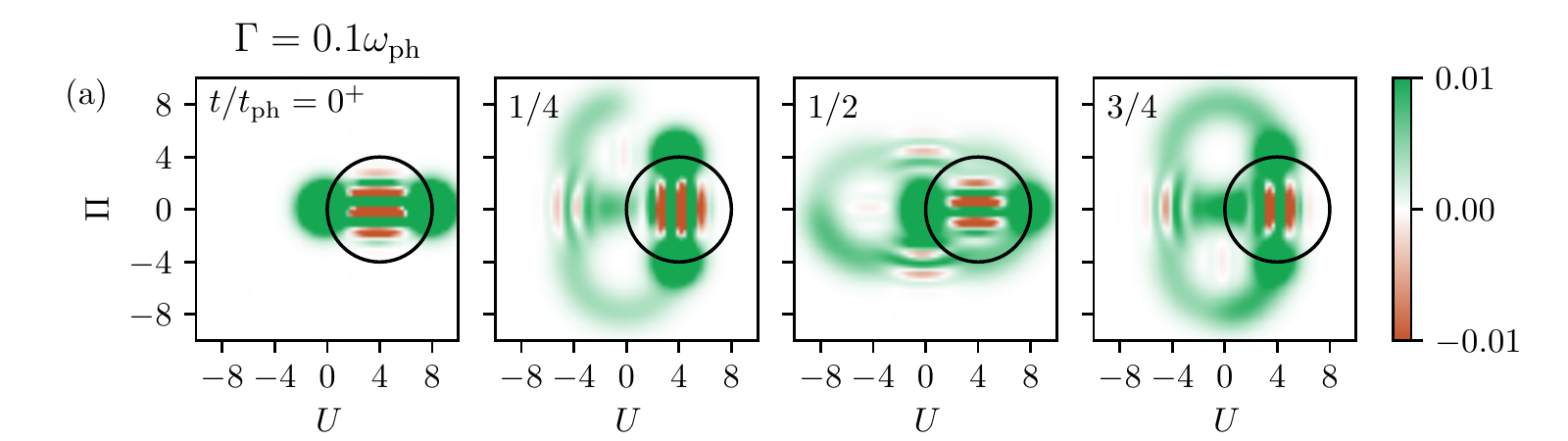}\\
		\includegraphics[width=\textwidth]{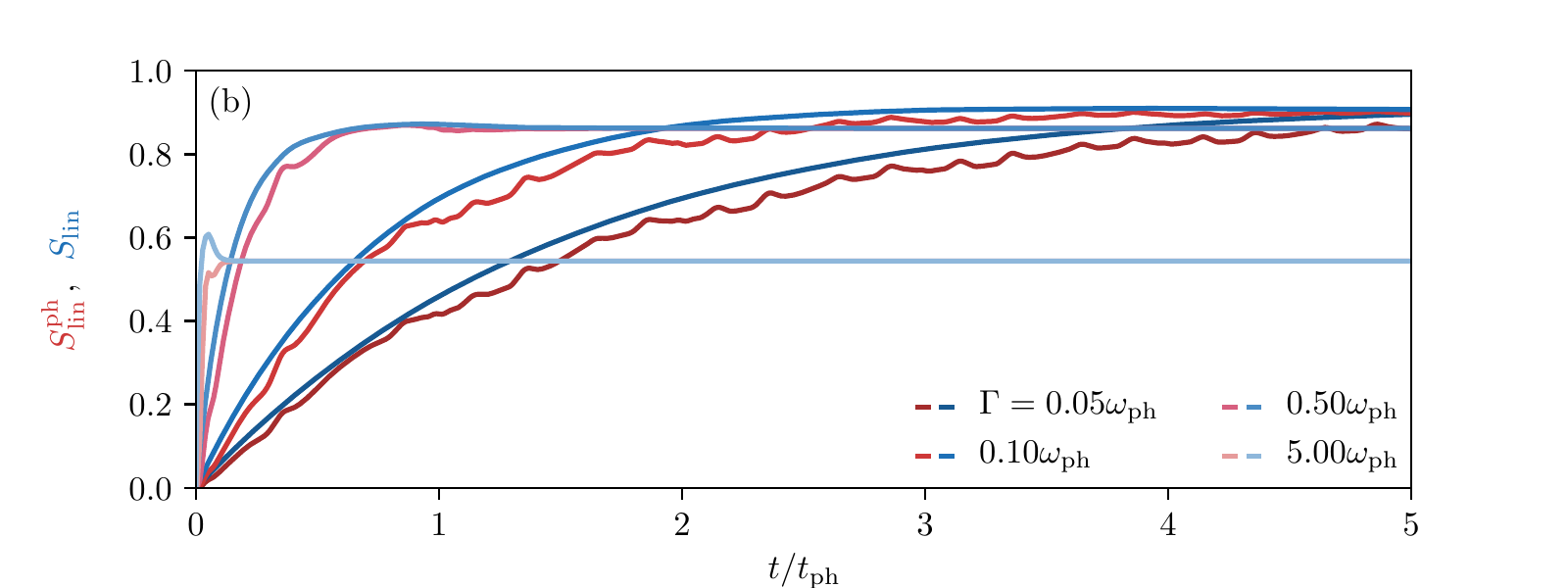}\\
		\includegraphics[width=\textwidth]{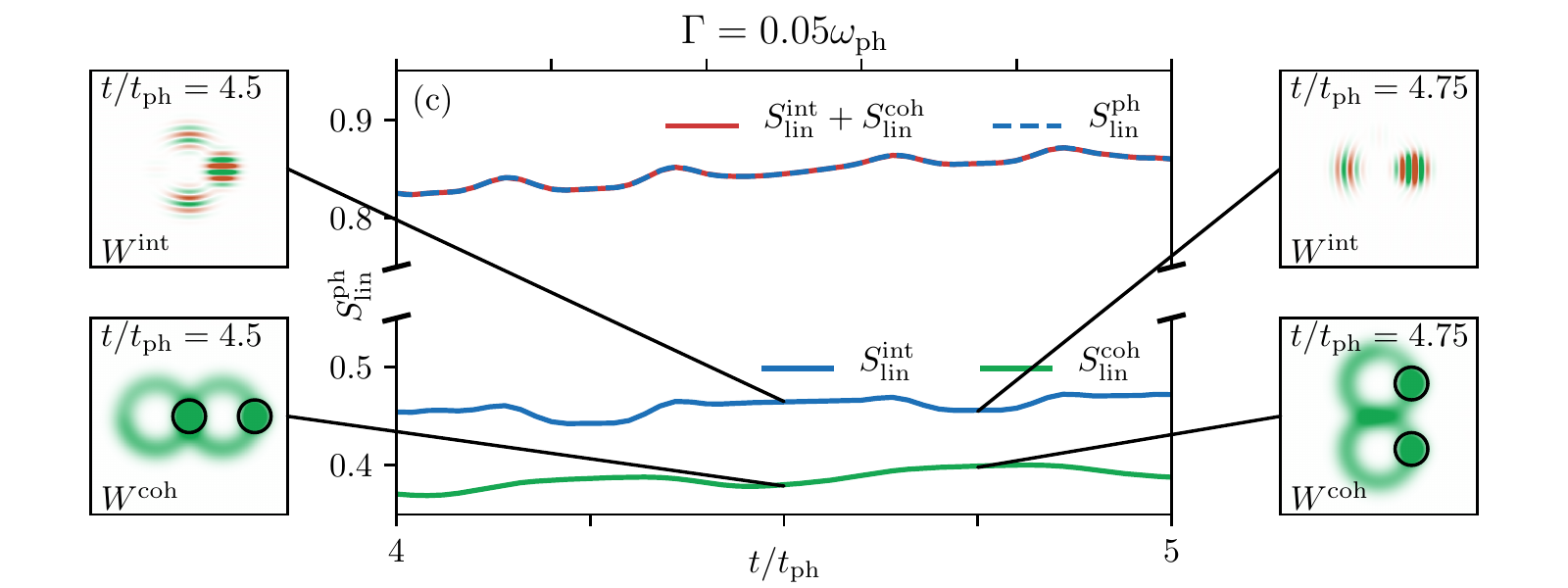}
	\caption{(\textbf{a}) Snapshots of the Wigner function during the decay into the ground state for a decay rate of $\Gamma=0.1\omega_{\rm ph}$. (\textbf{b}) Linear phonon entropies in red and full linear entropies in blue as functions of time for different decay rates. $\Gamma$ increases from bright to dark colors. (\textbf{c}) Zoom-in on one phonon period for $\Gamma = 0.05\omega_{\rm ph}$. The~coherent contribution $S_{\rm lin}^{\rm coh}$ is green, the~one from the interference $S_{\rm lin}^{\rm int}$ is blue and their sum dashed blue. Next to (\textbf{c}) are exemplary Wigner functions of the coherent part (bottom) and the interference (top) for the respective minima marked by black~lines.}\label{fig:decay_cat}
\end{figure}

To understand the origin of these dynamics in the phonon system, we take a closer look at the different parts of the Wigner function. According to Reference~\cite{reiter2011gene}, the~Wigner function of a cat state can be separated into
\begin{equation}
		W = W_{\rm coh} + W_{\rm int}\ ,
	\end{equation}
where $W_{\rm coh}$ describes the two coherent states that have been studied previously, and~$W_{\rm int}$ is the interference showing up as striped structure in phase space. Under~the assumption that the phonon coupling strength, $\gamma$ is large enough such that the different parts of the Wigner function do not significantly overlap in phase space; the linear entropy can also be separated into two contributions, $S_{\rm lin}^{\rm coh}$ and $S_{\rm lin}^{\rm int}$, calculated from the respective contributions of the Wigner function, and~we obtain
\begin{equation}
		S_{\rm lin}^{\rm ph} \approx S_{\rm lin}^{\rm coh} + S_{\rm lin}^{\rm int}\ .
	\end{equation}
In Figure~\ref{fig:decay_cat}c, we show the different entropies for a short time window $4\leqslant t/t_{\rm ph}\leqslant 5$. This already clarifies the picture a bit. First of all, we find that $S_{\rm lin}^{\rm coh}$ (green) and $S_{\rm lin}^{\rm int}$ (blue) are approximately of the same size and the sum of the two parts, shown in dashed blue, agrees perfectly with the full linear phonon entropy~(red line). The~coherent part $S_{\rm lin}^{\rm coh}$ has reduced values at times $t/t_{\rm ph}=n/2$, and~$S_{\rm lin}^{\rm int}$ develops minima exactly between those times, i.e.,~at $t/t_{\rm ph}=(2n+1)/4$. Because~the shapes of the minima in the two contributions are not the same, the~sum appears quite~involved.

After identifying the different dynamics, we have to understand their origin. Therefore, next to Figure~\ref{fig:decay_cat}c, we plot the Wigner functions $W_{\rm coh}$ and $W_{\rm int}$ for the respective minima, as~marked by the black lines. Starting with the coherent part at the bottom, we recognize that the situation is equivalent to the one in Figure~\ref{fig:1pulse_decay}b,c. The~entropy is always reduced when the rotating Wigner function of the coherent states in the excited state subspace (marked by black circles) overlaps with parts of the Wigner function in the ground state subspace. In~the right example, at~$t=4.75t_{\rm ph}$, the~two Gaussians are clearly separated from the decayed part in the ground state. In~the left one, at~$t=4.5t_{\rm ph}$, one of the coherent states overlaps with the touching point of the two circles that are in the ground state $\left| g\right>$. Because~we start with two coherent states in $\left| x\right>$, the~periodicity of the minima is half the phonon period. Moving on to the Wigner function of the interference contribution at the top, we only see the expected striped patterns. The~times where $S_{\rm lin}^{\rm int}$ is reduced agree with the times $t/t_{\rm ph}=(2n+1)/4$, where the interference terms that survive the decay process and remain also after the full decay are transferred into the ground state system (see discussion in Reference~\cite{hahn2019infl}). As~seen in the depicted Wigner function on the right, at~these times (e.g., $t=4.75t_{\rm ph}$) in Figure~\ref{fig:decay_cat}c, one of the interference terms that were already transferred into $\left| g\right>$ perfectly overlaps with the single interference that is still in the excited state, resulting in the two separated structures in phase space. For~all other times, \mbox{three contributions} appear, two in $\left|g\right>$ and one in $\left|x\right>$, as~exemplarily shown on the left for $t=4.5t_{\rm ph}$. Thus, the~fundamental reason for the reduction of the entropy is that a mixture of two cat states is more pure than a mixture of three. The~perfectly overlapping interferences on the right make the corresponding Wigner function look more like a mixture of two states than of~three.

\subsubsection{Phonons Generated by a Decaying~TLS}
To conclude the discussion, we now take a look at the two pulse excitation discussed in Section~\ref{sec:2pulse_no_decay} and consider a non-vanishing decay rate of the TLS. In~Reference~\cite{hahn2019infl}, it was shown that the final phonon state is in good agreement with the eight-shaped Wigner function of the decayed single cat state previously analyzed. However, now the phonon generation leads to a statistical mixture of two cat states that are additionally smeared out in phase space. This is exemplarily shown by the Wigner functions in Figure~\ref{fig:2pulse_decay}a. Although~the quantum state of the system is more involved, the~linear entropy of the phonons depicted as red line in Figure~\ref{fig:2pulse_decay}b evolves in a well-structured manner. Especially after the excitation with the second pulse, marked by the dashed black line, the~dynamics resemble the ones in Figure~\ref{fig:2pulse}b with an additional increase of the curve according to the decay process. We find the same broad depressions for half periods and stronger oscillating ones for full periods. The~small additional entropy reductions discussed in Figure~\ref{fig:decay_cat} are also found here, as~shown by the zoom-in in Figure~\ref{fig:2pulse}c. However, compared to the effects of the two overlapping cat states, as~previously mentioned, they almost disappear. The~linear entropy of the full system shown as a blue line in Figure~\ref{fig:2pulse}b is always smaller than the phonon part and grows smoothly. This is in agreement with the situation without any decay, where the entropy of the full system was always zero while the phonon part was~non-vanishing.

\begin{figure}[H]
	\centering
		\includegraphics[width=\textwidth]{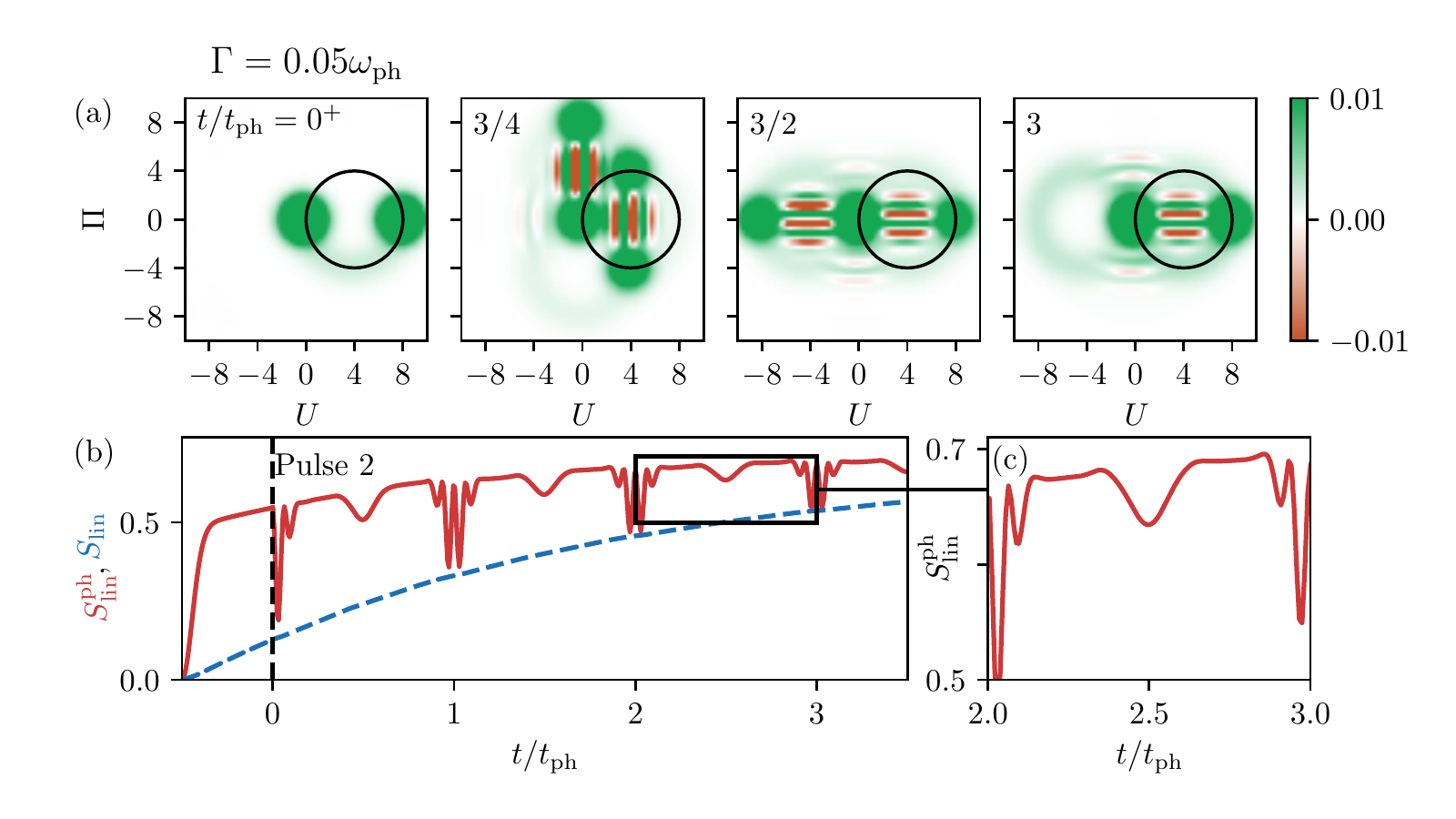}
	\caption{(\textbf{a}) Snapshots of the Wigner function's dynamics after the second pulse for $\Gamma=0.05\omega_{\rm ph}$. (\textbf{b})~Dynamics of the linear phonon entropy in red and the linear entropy of the full system in blue. (\textbf{c})~Zoom-in on the marked short time window from (\textbf{b}).}\label{fig:2pulse_decay}
\end{figure}

\section{Conclusions}
In summary, we analyzed the entropy dynamics of a single phonon mode coupled to an optically-driven TLS. We presented a theoretical framework that allowed us to calculate entropies of the different parts of the system when the quantum state of the entire system is pure and linear entropies when it is not pure. Additionally, the~concept of Wigner functions for the representation of phonon quantum states was used. We started our discussion with the most basic optical excitation, i.e.,~a single ultrafast pulse, that generated a mixture of two coherent states in the phonon system and assumed a non-decaying TLS. From~this, we further increased the complexity of the generated phonon state by including non-vanishing decay rates and two-pulse excitations of the TLS. This led to Wigner functions that smeared out in phase space and the generation of Schr\"odinger cat states, respectively. While the decay of the TLS, in~general, led to an increase of the system's entropy, the~complex dynamics of the phonon states resulted in temporally significant reductions of the phonon entropy. All these effects could be traced back to the purity of the quantum states and the entanglement between phonons and TLS. This extensive study on the phonon's entropy led to a thorough understanding of the fundamental interplay between the dynamics of the two separate parts and their combined influence on the quantum state~purity.

\vspace{6pt} 

\authorcontributions{conceptualization, T.H., D.W. and T.K.; methodology, T.H.; software, T.H.; validation, T.H., D.W. and T.K.; formal analysis, T.H.; investigation, T.H.; data curation, T.H.; writing--original draft preparation, D.W.; writing--review and editing, T.H., D.W. and T.K.; visualization, T.H. and D.W.; supervision, D.W. and T.K.; project administration, T.K. All authors have read and agreed to the published version of the manuscript.}

\funding{This research received no external funding.}

\conflictsofinterest{The authors declare no conflict of~interest.}

\reftitle{References}


\begin{thebibliography}{999}

\bibitem[Kondepudi and Prigogine(2014)]{kondepudi2014mode}
Kondepudi, D.; Prigogine, I.
\newblock {\em Modern Thermodynamics: From Heat Engines to Dissipative
  Structures}; John Wiley \& Sons: Berlin, Germany,  2014.

\bibitem[Gunzig \em{et~al.}(1987)Gunzig, Geheniau, and
  Prigogine]{gunzig1987entr}
Gunzig, E.; Geheniau, J.; Prigogine, I.
\newblock Entropy and cosmology.
\newblock {\em Nature} {\bf 1987}, {\em 330},~621--624.

\bibitem[la~Cruz-Dombriz and S{\'a}ez-G{\'o}mez(2012)]{la2012blac}
la~Cruz-Dombriz, D.; S{\'a}ez-G{\'o}mez, D.
\newblock Black holes, cosmological solutions, future singularities, and their
  thermodynamical properties in modified gravity theories.
\newblock {\em Entropy} {\bf 2012}, {\em 14},~1717--1770.

\bibitem[Toda \em{et~al.}(1992)Toda, Kubo, and Sait{\=o}]{toda1992equi}
Toda, M.; Kubo, R.; Sait{\=o}, N.
\newblock {\em Equilibrium Statistical Mechanics}; Springer: Berlin,
  Germany,  1992; Volume~1.

\bibitem[Kubo \em{et~al.}(2012)Kubo, Toda, and Hashitsume]{kubo2012stat}
Kubo, R.; Toda, M.; Hashitsume, N.
\newblock {\em Statistical Physics II: Nonequilibrium Statistical Mechanics};
  Springer: Berlin, Germany,  2012; Volume~31, 

\bibitem[Breuer and Petruccione(2002)]{breuer2002theo}
Breuer, H.P.; Petruccione, F.
\newblock {\em The Theory of Open Quantum Systems}; Oxford University Press:
  Oxford, UK,  2002.

\bibitem[Eisert \em{et~al.}(2010)Eisert, Cramer, and Plenio]{eisert2010coll}
Eisert, J.; Cramer, M.; Plenio, M.B.
\newblock {Colloquium: Area laws for the entanglement entropy}.
\newblock {\em Rev. Mod. Phys.} {\bf 2010}, {\em 82},~277.

\bibitem[Shannon(1948)]{shannon1948math}
Shannon, C.E.
\newblock A mathematical theory of communication.
\newblock {\em Bell Labs Tech. J.} {\bf 1948}, {\em 27},~379--423.

\bibitem[Liang \em{et~al.}(2006)Liang, Shi, Li, and Wierman]{liang2006info}
Liang, J.; Shi, Z.; Li, D.; Wierman, M.J.
\newblock Information entropy, rough entropy and knowledge granulation in
  incomplete information systems.
\newblock {\em Int. J. Gen. Syst.} {\bf 2006}, {\em 35},~641--654.

\bibitem[Nielsen and Chuang(2002)]{nielsen2002quan}
Nielsen, M.A.; Chuang, I.
\newblock {\em Quantum Computation and Quantum Information}; American
  Association of Physics Teachers: College Park, MD, USA,  2002.

\bibitem[Hofheinz \em{et~al.}(2009)Hofheinz, Wang, Ansmann, Bialczak, Lucero,
  Neeley, O'connell, Sank, Wenner, Martinis, and Cleland]{hofheinz2009synt}
Hofheinz, M.; Wang, H.; Ansmann, M.; Bialczak, R.C.; Lucero, E.; Neeley, M.;
  O'connell, A.D.; Sank, D.; Wenner, J.; Martinis, J.M.; et al.
\newblock Synthesizing arbitrary quantum states in a superconducting resonator.
\newblock {\em Nature} {\bf 2009}, {\em 459},~546.

\bibitem[O’Connell \em{et~al.}(2010)O’Connell, Hofheinz, Ansmann, Bialczak,
  Lenander, Lucero, Neeley, Sank, Wang, Weides, Wenner, Martinis, and
  Cleland]{o2010qua}
O’Connell, A.D.; Hofheinz, M.; Ansmann, M.; Bialczak, R.C.; Lenander, M.;
  Lucero, E.; Neeley, M.; Sank, D.; Wang, H.; Weides, M.; et al.
\newblock Quantum ground state and single-phonon control of a mechanical
  resonator.
\newblock {\em Nature} {\bf 2010}, {\em 464},~697--703.

\bibitem[Reiter \em{et~al.}(2011)Reiter, Wigger, Axt, and Kuhn]{reiter2011gene}
Reiter, D.E.; Wigger, D.; Axt, V.M.; Kuhn, T.
\newblock Generation and dynamics of phononic cat states after optical
  excitation of a quantum dot.
\newblock {\em Phys. Rev. B} {\bf 2011}, {\em 84},~195327.

\bibitem[Satzinger \em{et~al.}(2018)Satzinger, Zhong, Chang, Peairs, Bienfait,
  Chou, Cleland, Conner, Dumur, Grebel, Gutierrez, November, Povey, Whiteley,
  Awschalom, Schuster, and Cleland]{satzinger2018quan}
Satzinger, K.J.; Zhong, Y.P.; Chang, H.S.; Peairs, G.A.; Bienfait, A.; Chou,
  M.H.; Cleland, A.Y.; Conner, C.R.; Dumur, {\'E}.; Grebel, J.; et al.
\newblock Quantum control of surface acoustic-wave phonons.
\newblock {\em Nature} {\bf 2018}, {\em 563},~661--665.

\bibitem[Mahan(1981)]{mahan1981many}
Mahan, G.D.
\newblock {\em Many-Particle Physics}; Plenum Press: New York, NY, USA,  1981.

\bibitem[Zrenner \em{et~al.}(2002)Zrenner, Beham, Stufler, Findeis, Bichler,
  and Abstreiter]{zrenner2002cohe}
Zrenner, A.; Beham, E.; Stufler, S.; Findeis, F.; Bichler, M.; Abstreiter, G.
\newblock Coherent properties of a two-level system based on a quantum-dot
  photodiode.
\newblock {\em Nature} {\bf 2002}, {\em 418},~612.

\bibitem[Aharonovich \em{et~al.}(2011)Aharonovich, Castelletto, Simpson, Su,
  Greentree, and Prawer]{aharonovich2011diam}
Aharonovich, I.; Castelletto, S.; Simpson, D.A.; Su, C.H.; Greentree, A.D.;
  Prawer, S.
\newblock Diamond-based single-photon emitters.
\newblock {\em Prog. Phys.} {\bf 2011}, {\em 74},~076501.

\bibitem[Wigger \em{et~al.}(2020)Wigger, Karakhanyan, Schneider, Kamp,
  H{\"o}fling, Machnikowski, Kuhn, and Kasprzak]{wigger2019phon}
Wigger, D.; Karakhanyan, V.; Schneider, C.; Kamp, M.; H{\"o}fling, S.;
  Machnikowski, P.; Kuhn, T.; Kasprzak, J.
\newblock Acoustic phonon sideband dynamics during polaron formation in a
  single quantum dot.
\newblock {\em Opt. Lett.} {\bf 2020}, {\em 45},~919--922.

\bibitem[Roca \em{et~al.}(1994)Roca, Trallero-Giner, and Cardona]{roca1994pola}
Roca, E.; Trallero-Giner, C.; Cardona, M.
\newblock Polar optical vibrational modes in quantum dots.
\newblock {\em Phys. Rev. B} {\bf 1994}, {\em 49},~13704.

\bibitem[Gali \em{et~al.}(2011)Gali, Simon, and Lowther]{gali2011ab}
Gali, A.; Simon, T.; Lowther, J.E.
\newblock An ab~initio study of local vibration modes of the nitrogen-vacancy
  center in diamond.
\newblock {\em New J. Phys.} {\bf 2011}, {\em 13},~025016.

\bibitem[Debald \em{et~al.}(2002)Debald, Brandes, and Kramer]{debald2002cont}
Debald, S.; Brandes, T.; Kramer, B.
\newblock Control of dephasing and phonon emission in coupled quantum dots.
\newblock {\em Phys. Rev. B} {\bf 2002}, {\em 66},~041301.

\bibitem[Munsch \em{et~al.}(2017)Munsch, Kuhlmann, Cadeddu, G{\'e}rard,
  Claudon, Poggio, and Warburton]{munsch2017reso}
Munsch, M.; Kuhlmann, A.V.; Cadeddu, D.; G{\'e}rard, J.M.; Claudon, J.; Poggio,
  M.; Warburton, R.J.
\newblock Resonant~driving of a single photon emitter embedded in a mechanical
  oscillator.
\newblock {\em Nat. Commun.} {\bf 2017}, {\em 8},~76.

\bibitem[Ferry(1991)]{ferry1991semiconductors}
Ferry, D.K.
\newblock {\em Semiconductors}; Macmillan: New York, NY, USA, 1991.

\bibitem[Munn and Silbey(1978)]{munn1978theo}
Munn, R.W.; Silbey, R.
\newblock Theory of exciton transport with quadratic exciton--phonon coupling.
\newblock {\em J. Chem. Phys.} {\bf 1978}, {\em 68},~2439--2450.

\bibitem[Muljarov and Zimmermann(2004)]{muljarov2004deph}
Muljarov, E.A.; Zimmermann, R.
\newblock Dephasing in quantum dots: Quadratic coupling to acoustic phonons.
\newblock {\em Phys. Rev. Lett.} {\bf 2004}, {\em 93},~237401.

\bibitem[Machnikowski(2006)]{machnikowski2006chan}
Machnikowski, P.
\newblock Change of decoherence scenario and appearance of localization due to
  reservoir anharmonicity.
\newblock {\em Phys. Rev. Lett.} {\bf 2006}, {\em 96},~140405.

\bibitem[Chenu \em{et~al.}(2019)Chenu, Shiau, and Combescot]{chenu2019two}
Chenu, A.; Shiau, S.Y.; Combescot, M.
\newblock {Two-level system coupled to phonons: Full analytical solution}.
\newblock {\em Phys.~Rev. B} {\bf 2019}, {\em 99},~014302.

\bibitem[Duke and Mahan(1965)]{duke1965phon}
Duke, C.B.; Mahan, G.D.
\newblock {Phonon-broadened impurity spectra. I. Density of states}.
\newblock {\em Phys. Rev.} {\bf 1965}, {\em 139},~A1965.

\bibitem[Stock \em{et~al.}(2011)Stock, Dachner, Warming, Schliwa, Lochmann,
  Hoffmann, Toropov, Bakarov, Derebezov, Richter, A., Knorr, and
  Bimberg]{stock2011acou}
Stock, E.; Dachner, M.R.; Warming, T.; Schliwa, A.; Lochmann, A.; Hoffmann, A.;
  Toropov, A.I.; Bakarov,~A.K.; Derebezov, I.A.; Richter, M.; et al.
\newblock Acoustic and optical phonon scattering in a single In (Ga) As quantum
  dot.
\newblock {\em Phys. Rev. B} {\bf 2011}, {\em 83},~041304.

\bibitem[Wigger \em{et~al.}(2018)Wigger, Schneider, Gerhardt, Kamp,
  H{\"o}fling, Kuhn, and Kasprzak]{wigger2018rabi}
Wigger, D.; Schneider, C.; Gerhardt, S.; Kamp, M.; H{\"o}fling, S.; Kuhn, T.;
  Kasprzak, J.
\newblock Rabi oscillations of a quantum dot exciton coupled to acoustic
  phonons: Coherence and population readout.
\newblock {\em Optica} {\bf 2018}, {\em 5},~1442--1450.

\bibitem[Ramsay \em{et~al.}(2010)Ramsay, Godden, Boyle, Gauger, Nazir, Lovett,
  Fox, and Skolnick]{ramsay2010phon}
Ramsay, A.J.; Godden, T.M.; Boyle, S.J.; Gauger, E.M.; Nazir, A.; Lovett, B.W.;
  Fox, A.M.; Skolnick, M.S.
\newblock {Phonon-induced Rabi-frequency renormalization of optically driven
  single InGaAs/GaAs quantum dots}.
\newblock {\em Phys. Rev. Lett.} {\bf 2010}, {\em 105},~177402.

\bibitem[Hahn \em{et~al.}(2019)Hahn, Groll, Kuhn, and Wigger]{hahn2019infl}
Hahn, T.; Groll, D.; Kuhn, T.; Wigger, D.
\newblock Influence of excited state decay and dephasing on phonon quantum
  state preparation.
\newblock {\em Phys. Rev. B} {\bf 2019}, {\em 100},~024306.

\bibitem[Auffeves and Richard(2014)]{auffeves2014opti}
Auffeves, A.; Richard, M.
\newblock Optical driving of macroscopic mechanical motion by a single
  two-level system.
\newblock {\em Phys. Rev. A} {\bf 2014}, {\em 90},~023818.

\bibitem[Schleich(2011)]{schleich2011quan}
Schleich, W.P.
\newblock {\em Quantum Optics in Phase Space}; John Wiley \& Sons: Berlin,
  Germany,  2011.

\bibitem[Gerry and Knight(2005)]{gerry2005intr}
Gerry, C.; Knight, P.L.
\newblock {\em Introductory Quantum Optics}; Cambridge University Press:
  Cambridge, UK,  2005.

\bibitem[Von~Neumann(1996)]{von1996math}
Von~Neumann, J.
\newblock {\em {Mathematische Grundlagen der Quantenmechanik}};
  Springer: Berlin, Germany,  1996;  Volume~2.

\bibitem[Wehrl(1978)]{wehrl1978gene}
Wehrl, A.
\newblock General properties of entropy.
\newblock {\em Rev. Mod. Phys.} {\bf 1978}, {\em 50},~221.

\bibitem[Manfredi and Feix(2000)]{manfredi2000entr}
Manfredi, G.; Feix, M.R.
\newblock Entropy and Wigner functions.
\newblock {\em Phys. Rev. E} {\bf 2000}, {\em 62},~4665.

\bibitem[Brune \em{et~al.}(1992)Brune, Haroche, Raimond, Davidovich, and
  Zagury]{brune1992mani}
Brune, M.; Haroche, S.; Raimond, J.M.; Davidovich, L.; Zagury, N.
\newblock Manipulation of photons in a cavity by dispersive atom-field
  coupling: Quantum-nondemolition measurements and generation of
  ‘‘Schr{\"o}dinger cat’’states.
\newblock {\em Phys. Rev. A} {\bf 1992}, {\em 45},~5193.

\bibitem[Ourjoumtsev \em{et~al.}(2007)Ourjoumtsev, Jeong, Tualle-Brouri, and
  Grangier]{ourjoumtsev2007gene}
Ourjoumtsev, A.; Jeong, H.; Tualle-Brouri, R.; Grangier, P.
\newblock Generation of optical ‘Schr{\"o}dinger cats’ from photon number
  states.
\newblock {\em Nature} {\bf 2007}, {\em 448},~784--786.

\bibitem[Deleglise \em{et~al.}(2008)Deleglise, Dotsenko, Sayrin, Bernu, Brune,
  Raimond, and Haroche]{deleglise2008reco}
Deleglise, S.; Dotsenko, I.; Sayrin, C.; Bernu, J.; Brune, M.; Raimond, J.M.;
  Haroche, S.
\newblock Reconstruction of non-classical cavity field states with snapshots of
  their decoherence.
\newblock {\em Nature} {\bf 2008}, {\em 455},~510.

\end{thebibliography}
\end{document}